\documentclass[reprint]{JASA}
\usepackage{lmodern}

\begin{document}

\title[]{Increasing ultrasound field-of-view with reduced element count arrays containing large elements}

\author{Mick Gardner}
\email{mickhg2@illinois.edu}

\author{Rita J. Miller}
\author{Michael L. Oelze}

\affiliation{Electrical and Computer Engineering,  University of Illinois Urbana-Champaign, Urbana, IL 61801, USA}	

\begin{abstract}
Several applications of medical ultrasound can benefit from a larger field of view (FOV). This study is aimed at increasing the FOV of linear array probes by increasing the element width. Coupled elements were used to imitate a larger element width. Through Fourier analysis, theoretical pressure amplitudes, and bandwidth estimates, coupled elements are shown to be close approximations of large elements. The effects of coupling on resolution, contrast, and speckle signal-to-noise ratio are investigated through phantom images and in-vivo images of a rabbit tumor reconstructed with plane-wave compounding. Furthermore, a positioning system was used to acquire data from a virtual large aperture with 120 mm FOV and 128 elements, collected in sections with a single probe. The Null Subtraction Imaging (NSI), Sign Coherence Factor (SCF), and Minimum Variance (MV) beamformers are compared for regaining resolution lost by an increased F-number. The NSI beamformer decreased Full-Width at Half-Max (FWHM) estimates of wire targets by 79\% with coupling by 2 compared to uncoupled DAS. The MV beamformer was best for maintaining speckle statistics while improving resolution. Our results demonstrate how increased element width can increase FOV with no increase to element count. 
\end{abstract}

\maketitle

\section{Introduction}
Increasing the field of view (FOV) of ultrasound is of major interest in the medical ultrasound community for applications such as abdominal \cite{kim_extended_2003}, muscle \cite{noorkoiv_assessment_2010}, spine \cite{huang_scoliotic_2019}, and vascular imaging \cite{kang_wide_2020,wang_wide_2022}. Many structures and organs in the human body are too large to fit in a single image with current ultrasound probes. This is especially true for the case of 3D ultrasound imaging using a 2D matrix array, where the field-of-view is often severely limited by a small probe footprint with an extremely high channel count. Therefore, there is a need for an increased field-of-view to capture a full region of interest. Several approaches exist to increasing the FOV. 

One such approach is to add more elements to the array \cite{foiret_improving_2022,bottenus_impact_2020}. Adding more elements will increase the FOV by increasing the probe footprint, and can also provide benefits to resolution from an increased F-number. However, cost, data processing, and electrical connections to the back-end system all become a burden for very high channel count arrays. Two approaches exist for handling high channel counts: multiplexing and micro-beamforming. With multiplexing, the system switches between subsections of the array for different transmit/receive events, so a high channel count probe can be connected to a lower input count system \cite{yu_design_2020}. The primary trade-off with multiplexing is the frame rate. For a probe with a 4 x 1 multiplexer, anywhere between 4 - 16 transmission events are needed for a single acquisition (e.g. plane-wave angle) \cite{chavignon_3d_2022}. With micro-beamforming, elements are grouped into patches, beamforming within the patches is done on ASICs inside the transducer handle, then beamforming across patches is done by the back-end system \cite{savord_fully_2003}. This way the probe only needs one connection per patch back to the system. However, micro-beamformers are suboptimal for ultrafast imaging, due to quantization and approximation of the time delay calculations in the ASICs \cite{castrignano_impact_2025}.

Alternative approaches seek to increase the FOV without an increased element count. For example, convex and phased arrays use limited elements, but have FOVs that extend far beyond their footprint due to the convexity or beam steering. A challenge with these probes is maintaining resolution over the FOV with scan lines that spread out with depth \cite{kang_wide_2020}. Panoramic images, sometimes called extended FOV (EFOV) ultrasonography, presents another option for increasing the FOV \cite{kim_extended_2003,poon_three-dimensional_2006}. In this method, the probe is translated to multiple locations and images from the separate acquisitions are registered to create a larger combined image. However, this method could only create static images with the increased FOV, not real-time video data, because it requires acquisitions from multiple locations. Array sparsity, in other words breaking array periodicity, has also been investigated as a method for reducing the element count on both 1D and 2D arrays \cite{ramalli_design_2022,gavrilov_method_1997}. This is usually achieved by deactivating select elements from fully populated arrays. One issue with sparsity is decreased transmit power (lower SNR) due to low element count and small elements. Also, the break in periodicity creates higher side lobes, reducing contrast. 

Another approach to increasing the field of view without increased element count is the use of larger elements. Recently, for the case of 3D imaging, row-column arrays were introduced which consist of 2 orthogonal arrays of very large line elements accessed by row or column index \cite{jensen_anatomic_2022}. This has allowed for construction of much larger 2D arrays with wider FOV and reduced element count. Another group has tested large 2D apertures for ultrasound localization microscopy (ULM) that use large, circular elements \cite{favre_boosting_2022,favre_transcranial_2023}. In this approach, to address issues with resolution from an increased F-number of highly directive elements, diverging acoustic lenses were placed over individual elements to widen the directivity and regain resolution \cite{favre_boosting_2022}. However, lenses did not eliminate grating lobes. 

Rather than using lenses, a simpler solution could be to use adaptive or non-linear beamformers suitable for the task of regaining resolution lost by a narrowed directivity. The Null Subtraction Imaging (NSI) beamformer is an emerging non-linear beamforming technique that can greatly improve resolution by imaging with beam nulls instead of a main lobe \cite{agarwal_improving_2019,kou_high-resolution_2023}. NSI has also been used to mitigate grating lobes in rat tumor images \cite{kou_grating_2022} and on large pitch arrays \cite{gardner_grating_2024}. The minimum variance (MV) beamformer has been shown to greatly improve resolution in ultrasound B-mode images \cite{synnevag_adaptive_2007}. Lastly, the Sign Coherence Factor (SCF) beamformer was introduced as a computationally inexpensive version of Phase Coherence Factor (PCF) which was made to reduce grating lobes and can also improve resolution \cite{camacho_grating-lobes_2009,camacho_phase_2009}. 

This study is aimed at increasing the FOV of ultrasound by increasing the element width, then using an appropriate beamformer for regaining resolution and mitigating grating lobe artifacts. One potential application for larger 1D arrays is abdominal imaging, where organs such as the liver and kidneys are larger than most commercial probes. Another potential application is to increase the FOV for 3D imaging with large, square elements in a 2D matrix array. To test our approach, we examined larger elements on a 1D probe, using coupled elements to imitate a larger element width. The element coupling involved transmitting blocks of adjacent elements at the same time, then summing their receive radio-frequency (RF) channel data so that they acted as if they were one element. We have examined element directivities, element sensitivities, pressure outputs, and bandwidths to determine whether coupling makes a good approximation of a large element. We also performed phantom experiments to see the effects of coupled elements on image quality, as well as compared conventional delay-and-sum (DAS) beamforming with NSI, SCF, and MV beamformers for restoring resolution. To show how this approach can lead to larger apertures and FOVs, we also collected data from a virtual large aperture, with data collected in sections using a standard probe on a positioning system. Plane-wave transmissions were used for all experiments because they can provide much higher frame rates than traditional line-by-line scanning.

\section{Background Theory}
\subsection{Beam Patterns from Larger Elements}
Here we present a brief analysis on the effects of large elements on the beam pattern of a linear array. The beam pattern is the product of the directivity (sometimes called an element factor) and an array factor, as in
\begin{equation} \label{eq:beam_pattern}
    B(\theta) = H(\theta)G(\theta)
\end{equation}
where \(\theta\) is the angle of arrival of a signal, \(H(\theta)\) is the array factor, \(G(\theta)\) is the element factor, and \(B(\theta)\) is the final beam pattern.

The array factor for an unsteered point-source array at a single frequency is given by the following equation
\begin{equation} \label{eq:array_factor}
    H(\theta) = \frac{1}{N}\frac{sin\left( \frac{N}{2}kdsin(\theta) \right)}{sin\left( \frac{1}{2}kdsin(\theta) \right)}
\end{equation}
where \(\theta\) is the direction of arrival, \(N\) is the number of elements in the array, \(k\) is the wave-number, and \(d\) is the array pitch (spacing between elements). When \(d\) is greater than a wavelength, grating lobes will appear in the unsteered array factor, with locations found by setting the denominator of Eq. \ref{eq:array_factor} equal to zero and solving for \(\theta\):
\begin{equation} \label{eq:grating_lobe_locations}
    \theta = \pm sin^{-1}\left(\frac{m\lambda}{d} \right)
\end{equation}
where \(\lambda\) is the wavelength, and \(m\) is an index.

The element directivity for a single frequency is given by the formula
\begin{equation} \label{eq:directivity}
        G(\theta) = sinc \left(\frac{Wksin(\theta)}{2}\right) = \frac{sin(Wksin(\theta)/2)}{Wksin(\theta)/2}
\end{equation}
where \(W\) is the width of an element. When the width \(W\) is greater than a wavelength, nulls will appear in the directivity, with locations found by setting the numerator of Eq. \ref{eq:directivity} equal to zero and solving for \(\theta\):
\begin{equation} \label{eq:directivity_nulls}
    \theta = \pm \sin^{-1}\left(\frac{m\lambda}{W} \right).
\end{equation}

\begin{figure}
    \centering
    \includegraphics[width=\linewidth]{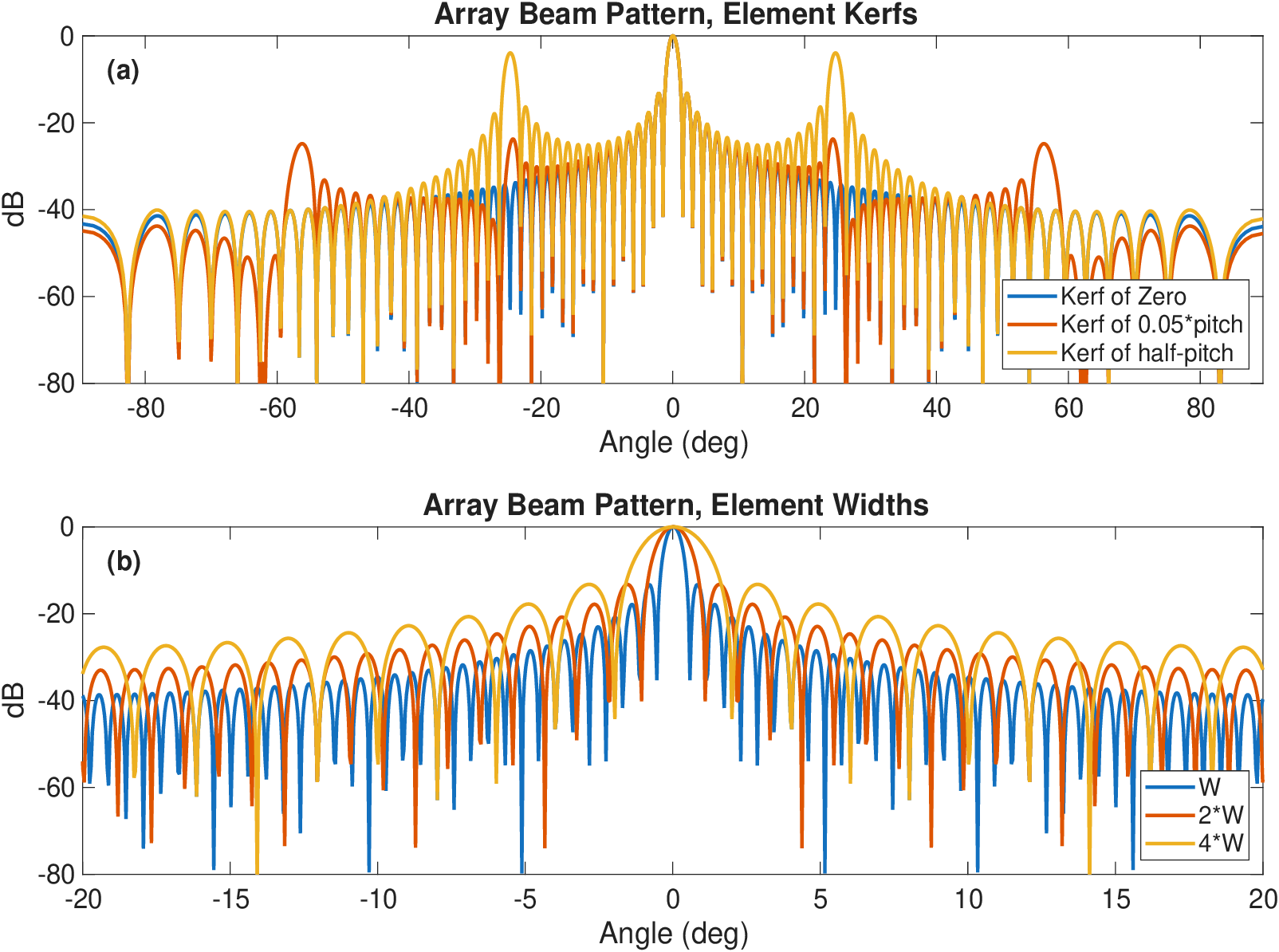}
    \caption{(a) Beam patterns for 16-element arrays with a pitch of 2.5 wavelengths and different kerfs. A kerf of 0 cancels all grating lobes, which means it is no longer an array but a single element. A kerf that is half the pitch cancels the second grating lobe but leaves the first grating lobes. The minimum kerf that can be manufactured is optimal to minimize all grating lobes. (b) Array beam patterns for different element widths. The larger element width/narrower directivity raises the array F-number, resulting in a wider main lobe.}
    \label{fig1:beams_kerfs_widths}
\end{figure}

With the multiplication in Eq. \ref{eq:beam_pattern}, it is desirable to line up element factor nulls with array factor grating lobes so that grating lobes are canceled. From Eqs. \ref{eq:grating_lobe_locations} and \ref{eq:directivity_nulls}, the width that lines up nulls and grating lobes is \(W = d\). This would represent a kerf of zero, where \(kerf = d - W\), representing the empty space between adjacent element edges. Of course, this case does not represent an array, but one single line element. For a linear ultrasound probe, the kerf is the width of the saw used to cut the piezoelectric material into separate elements. The kerf will slightly push out the nulls in the element factor, reintroducing grating lobes as displayed in Figure \ref{fig1:beams_kerfs_widths}a. The minimum kerf that can be manufactured is optimal for minimizing grating lobes, but grating lobe artifacts will still exist. As the pitch of the array is increased, more grating lobes will appear (Eq. \ref{eq:grating_lobe_locations}).

Also, larger elements make the element factor more narrow (Eq. \ref{eq:directivity}), which raises the achievable F-number of the array \cite{perrot_so_2021}. A higher F-number widens the main lobe, as displayed in Figure \ref{fig1:beams_kerfs_widths}b. Therefore, a loss in image resolution is expected with larger elements. These issues with resolution and grating lobes motivate the use of adaptive and non-linear beamformers which can improve resolution and reduce grating lobes \cite{kou_grating_2022,gardner_grating_2024,camacho_grating-lobes_2009,synnevag_adaptive_2007}. 

\subsection{Element Coupling to Approximate Large Elements}
\begin{figure}
    \centering
    \includegraphics[width=\linewidth]{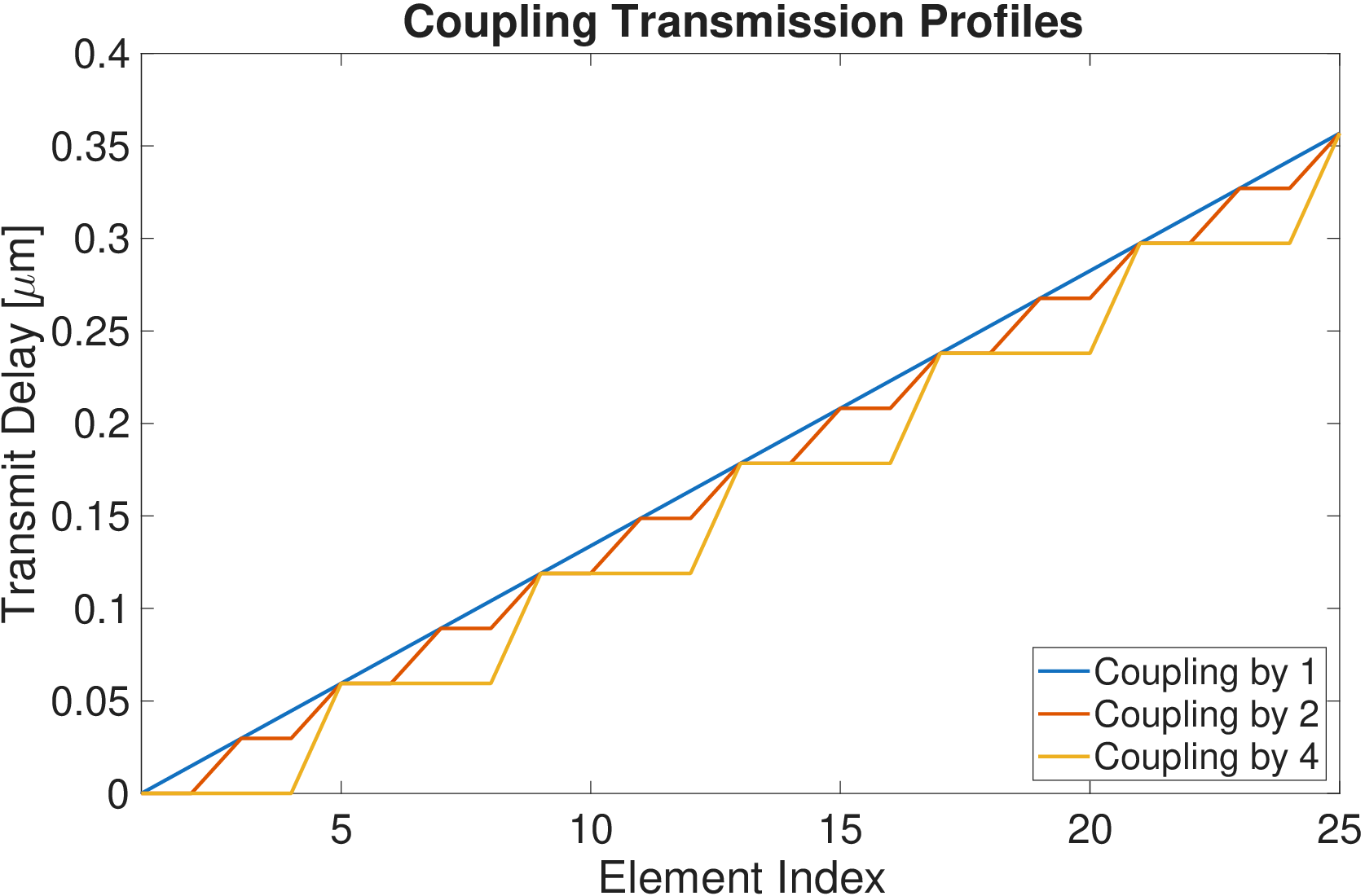}
    \caption{Example transmission profiles of uncoupled (blue), coupled by 2 (orange), and coupled by 4 (yellow) arrays for a positive plane-wave steering angle. Blocks of consecutive elements fire at the same time for coupling by 2 and 4, making the jagged steps in the transmit delay profile.}
    \label{fig7:coupled_transmit_profile}
\end{figure}

\begin{figure}
    \centering
    \includegraphics[width=\linewidth]{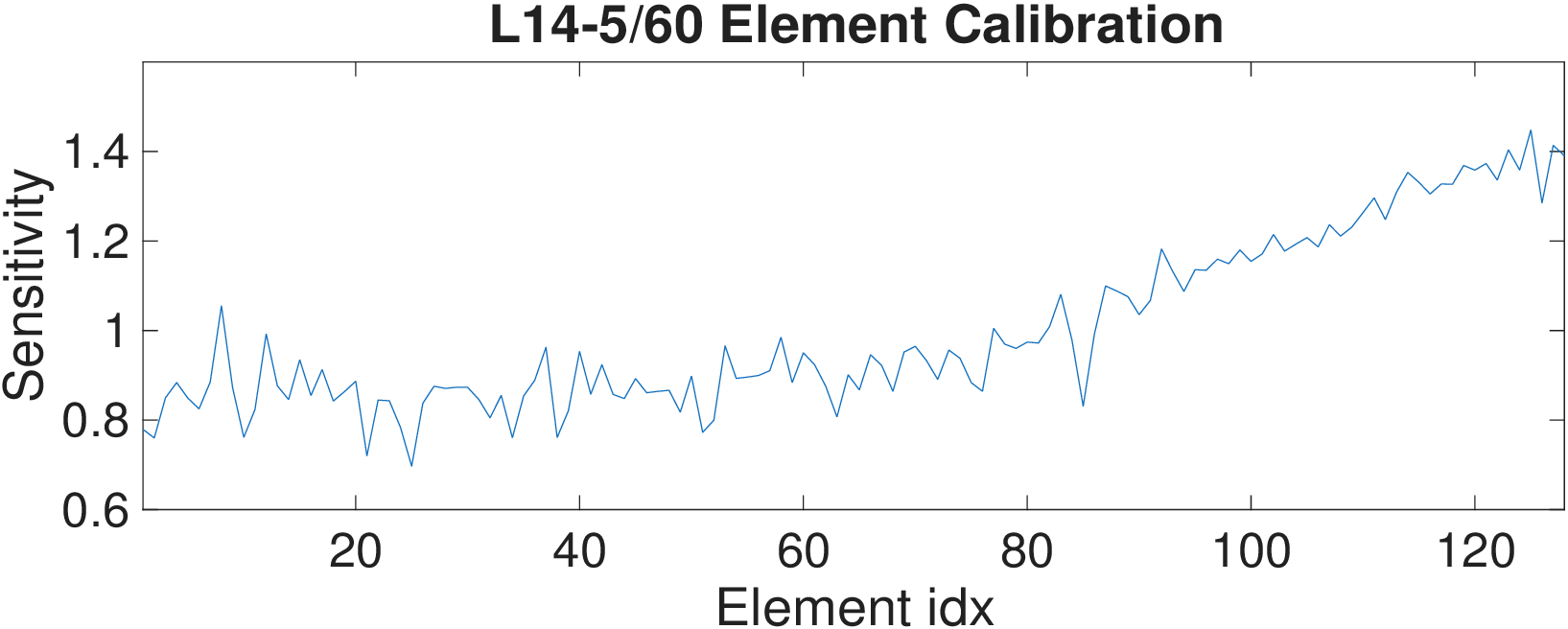}
    \caption{Estimated element sensitivities of the L14-5/60 probe.}
    \label{fig2:element_sensitivities}
\end{figure}

This study used coupled elements on an L14-5/60 probe (Ultrasonix, BC, Canada) to imitate larger elements. Element coupling was performed on both transmit and receive so that blocks of adjacent elements would act as if they were one large element. Elements were coupled by averaging their transmit delays and summing their received channel data. For a steered plane-wave, the transmit profiles of coupled and uncoupled elements looked something like Figure \ref{fig7:coupled_transmit_profile}, where blocks of adjacent elements fired at the same time. Then, the receive RF traces within each block were summed together without applying any time delays to create the coupled RF trace. The coupled elements still had kerf gaps in between, which a large element of the same width would not have. To determine how close an approximation coupling would make of a physically large element, we started by estimating the sensitivities of individual elements using an insertion loss measurement \cite{kou_high-resolution_2023}. Briefly, this measurement involves setting up the probe in a water tank with a planar reflector, firing and receiving from one element at a time, then calculating normalization weights based on the relative amplitude of received envelopes. The resulting estimated sensitivities are displayed in Figure \ref{fig2:element_sensitivities}. 

\begin{figure}
    \centering
    \includegraphics[width=\linewidth]{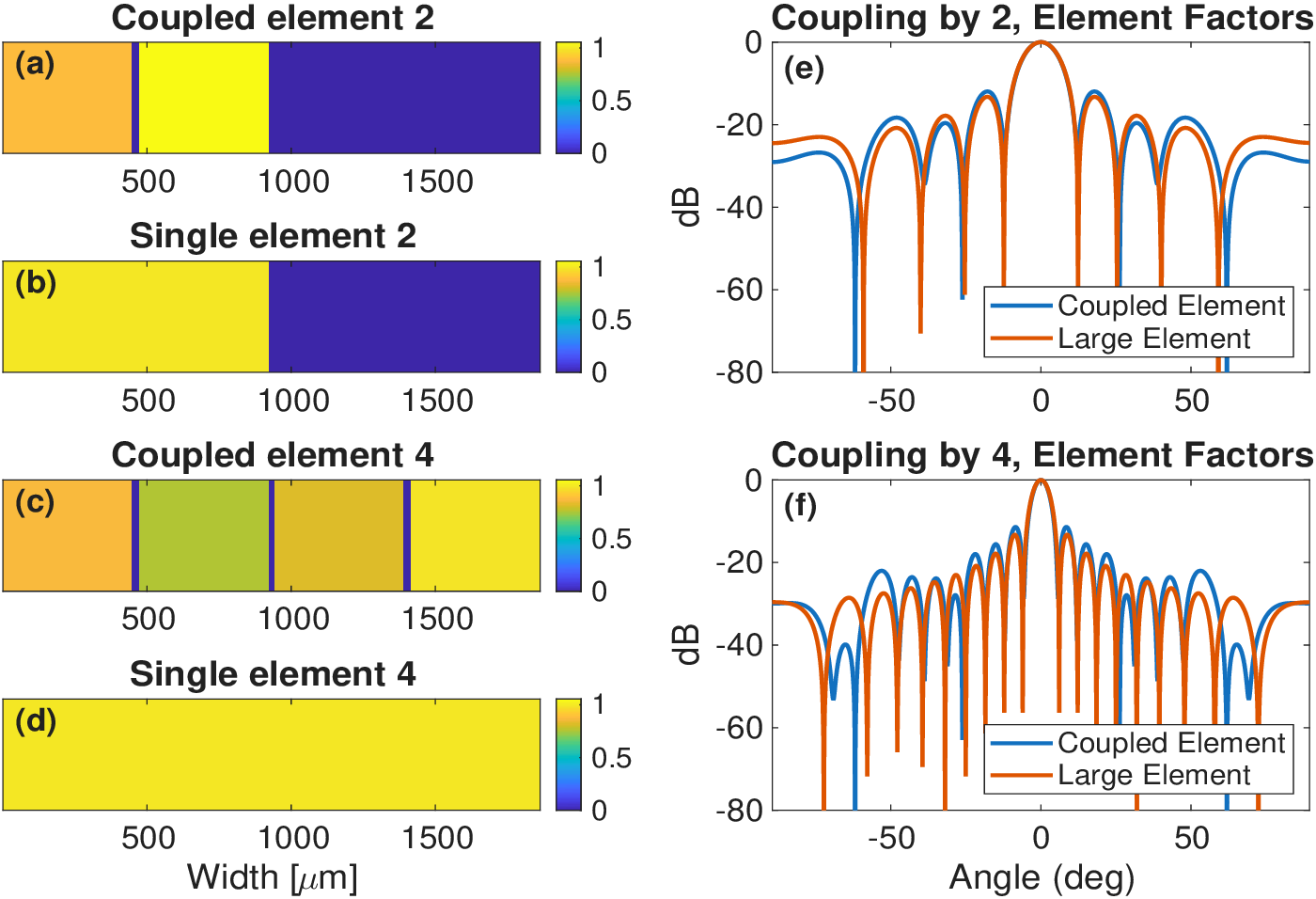}
    \caption{Directivities of coupled elements (with kerf gaps) compared to directivities of corresponding large elements. (a) Coupled element by 2 (top) and large element (bottom) with corresponding directivities in (b). (c) Coupled element by 4 (top) and large element (bottom) with corresponding directivities in (d). The directivities of the coupled elements match closely to those of the large elements, meaning coupled elements are a good approximation of a large element.}
    \label{fig3:coupled_directivities}
\end{figure}

Applying these sensitivity values to individual elements, we took the spatial Fourier transforms of coupled and large elements, displayed in Figure \ref{fig3:coupled_directivities}. The large elements are assumed to have a uniform gain of one across their face. In the Fourier transforms, only small differences in the side lobes are observed, while the main lobes are practically identical.

\begin{figure}
    \centering
    \includegraphics[width=\linewidth]{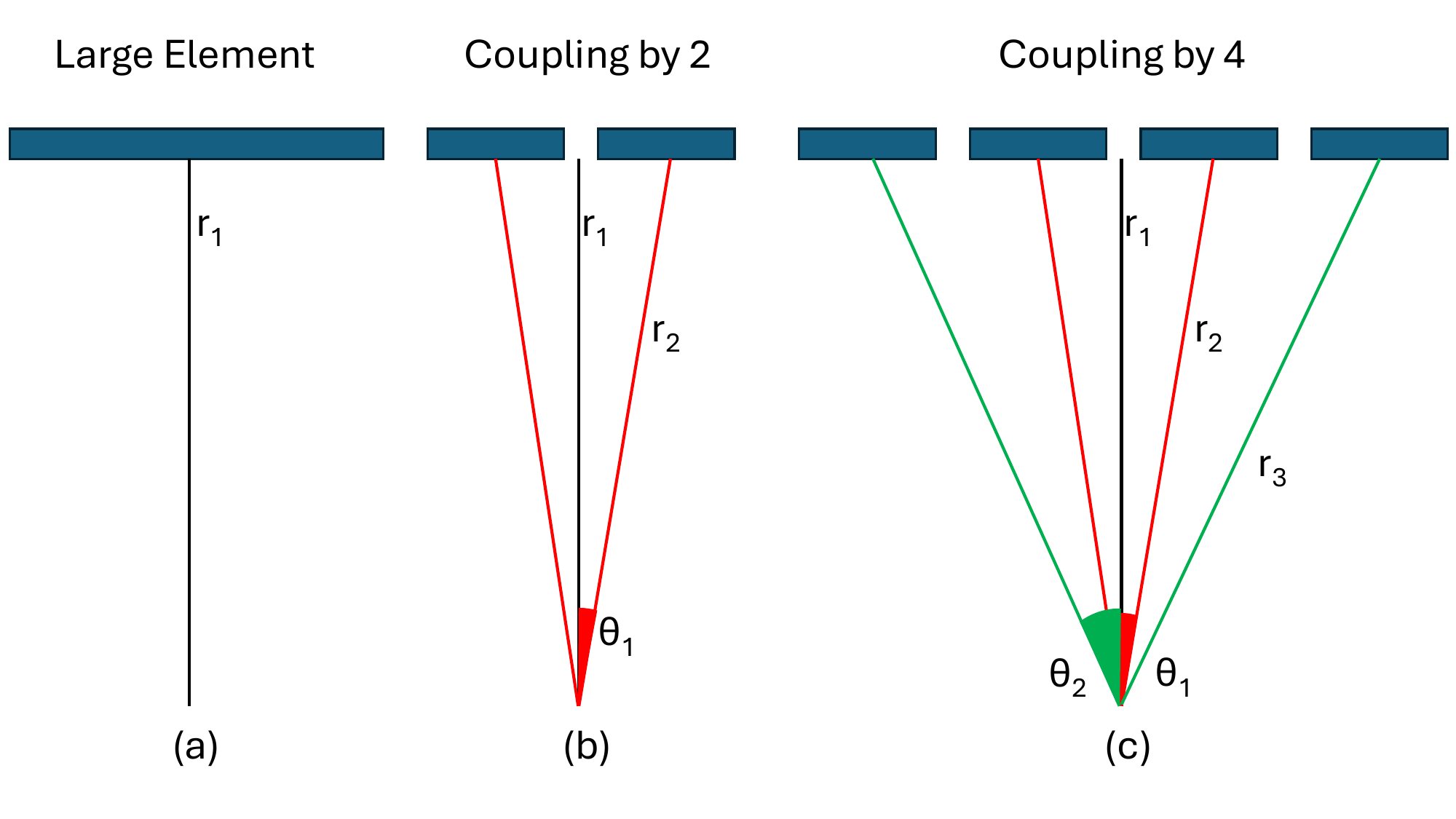}
    \caption{Illustration for axial pressure derivation for (a) large, (b) coupling by 2, and (c) coupling by 4.}
    \label{fig4:axial_pressure_diagram}
\end{figure}

\begin{figure}
    \centering
    \includegraphics[width=\linewidth]{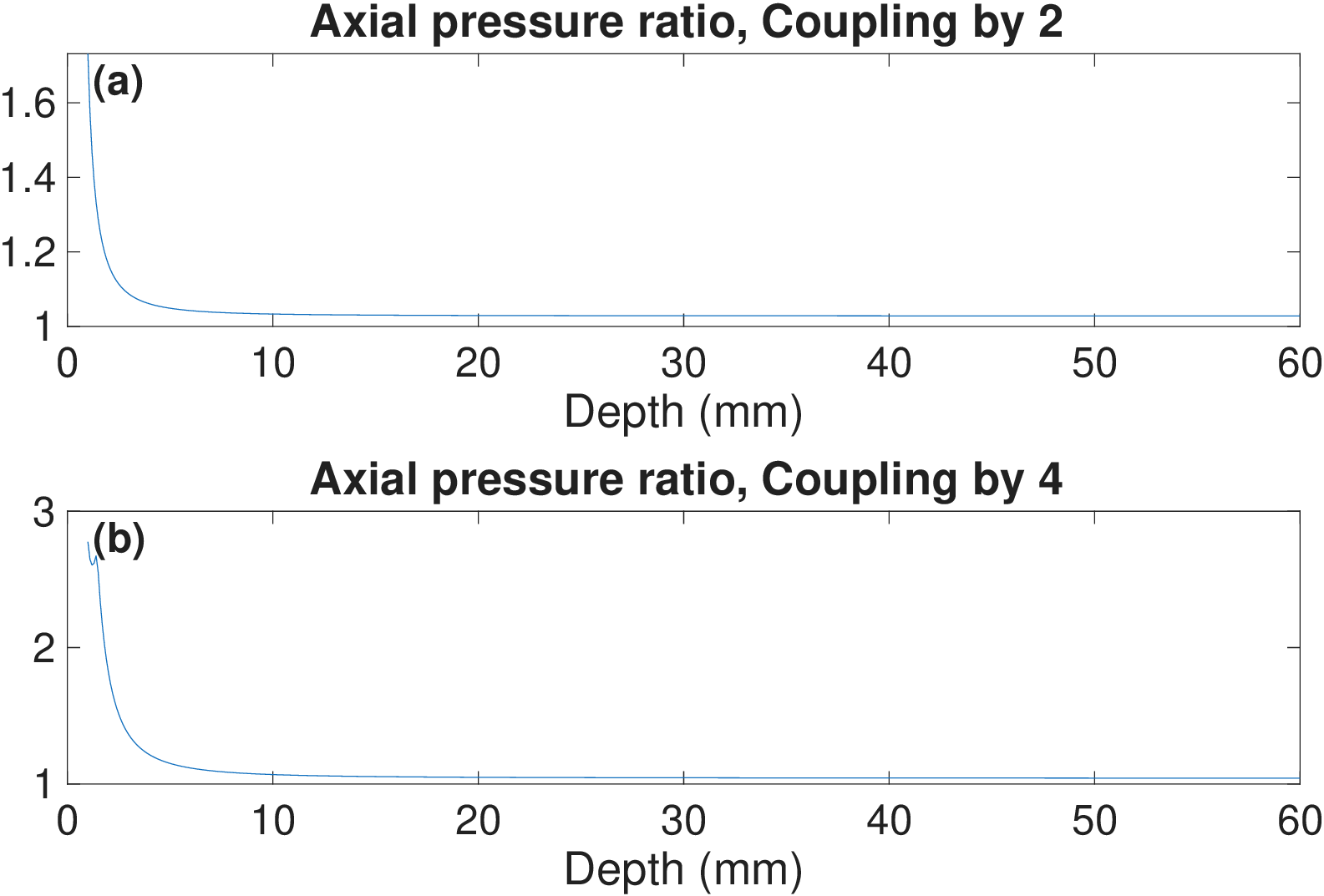}
    \caption{Ratios of axial pressure amplitudes for large versus coupled elements. (a) Ratio for coupling by 2. (b) Ratio for coupling by 4.}
    \label{fig5:pressure_ratios}
\end{figure}

Next, to examine the sensitivity differences, we derived the ratio of axial pressure amplitudes for large versus coupled elements. If we approximate the elements as cylindrical line elements, the axial pressure amplitude in the far field is given by \cite{kinsler_fundamentals_2012}
\begin{equation} \label{eq:axial_pressure}
    P_{large}(r) = \frac{1}{2} \rho_{0} c U_{0} \left(\frac{a}{r}\right) k L
\end{equation}
where \(\rho_{0}\) is the density of the medium, \(c\) is the sound speed, \(U_{0}\) is the particle velocity amplitude, \(a\) is the radius of the cylinder, \(r\) is the depth, \(k\) is the wavenumber, and \(L\) is the length of the source. For a large element, the axial pressure is simply given by Eq. \ref{eq:axial_pressure} with the appropriate length plugged in. For a coupled element, the axial pressure will be a sum of the individual element contributions. If we assume symmetry about the axis, the pressure is simply
\begin{equation}
    P_{coupled}(r) = 2 \left(\frac{1}{2} \rho_{0} c U_{0} \left(\frac{a}{r}\right) k W\right) G(\theta)
\end{equation}
where \(W\) is the width of individual elements, \(\theta\) is the angle between the center of an individual element and the axis, and \(G(\theta)\) is the element directivity (see Figure \ref{fig4:axial_pressure_diagram}). Taking the ratio cancels most of the terms and leaves the following expression for coupling by two,
\begin{equation}
    \frac{P_{large}}{P_{coupled}} = \frac{L/r_{1}}{2 W (1/r_{2}) G(\theta)}.
\end{equation}
A similar analysis for coupling by 4 creates a ratio of
\begin{equation}
    \frac{P_{large}}{P_{coupled}} = \frac{L/r_{1}}{2W\left[(1/r_3)G(\theta_2) + (1/r_2)G(\theta_1) \right]}.
\end{equation}
Plugging in element widths from the L14-5/60 probe, and plotting these expressions for \(r_{1}\) values (i.e. depths) of 1 mm to 60 mm yields the traces in Figure \ref{fig5:pressure_ratios}. For depths less than about 4 mm, where the plots spike upward, far-field approximations for axial pressure and directivity no longer hold, and this derivation is not accurate. For depths sufficiently large, above about 4 mm, the pressure from a large element is only 2\% higher for coupling by two, and about 4\% higher for coupling by four. Therefore, coupled elements transmit nearly the same amount of power into the medium, closely approximating the sensitivity one would expect for a large element.

\begin{figure}
    \centering
    \includegraphics[width=\linewidth]{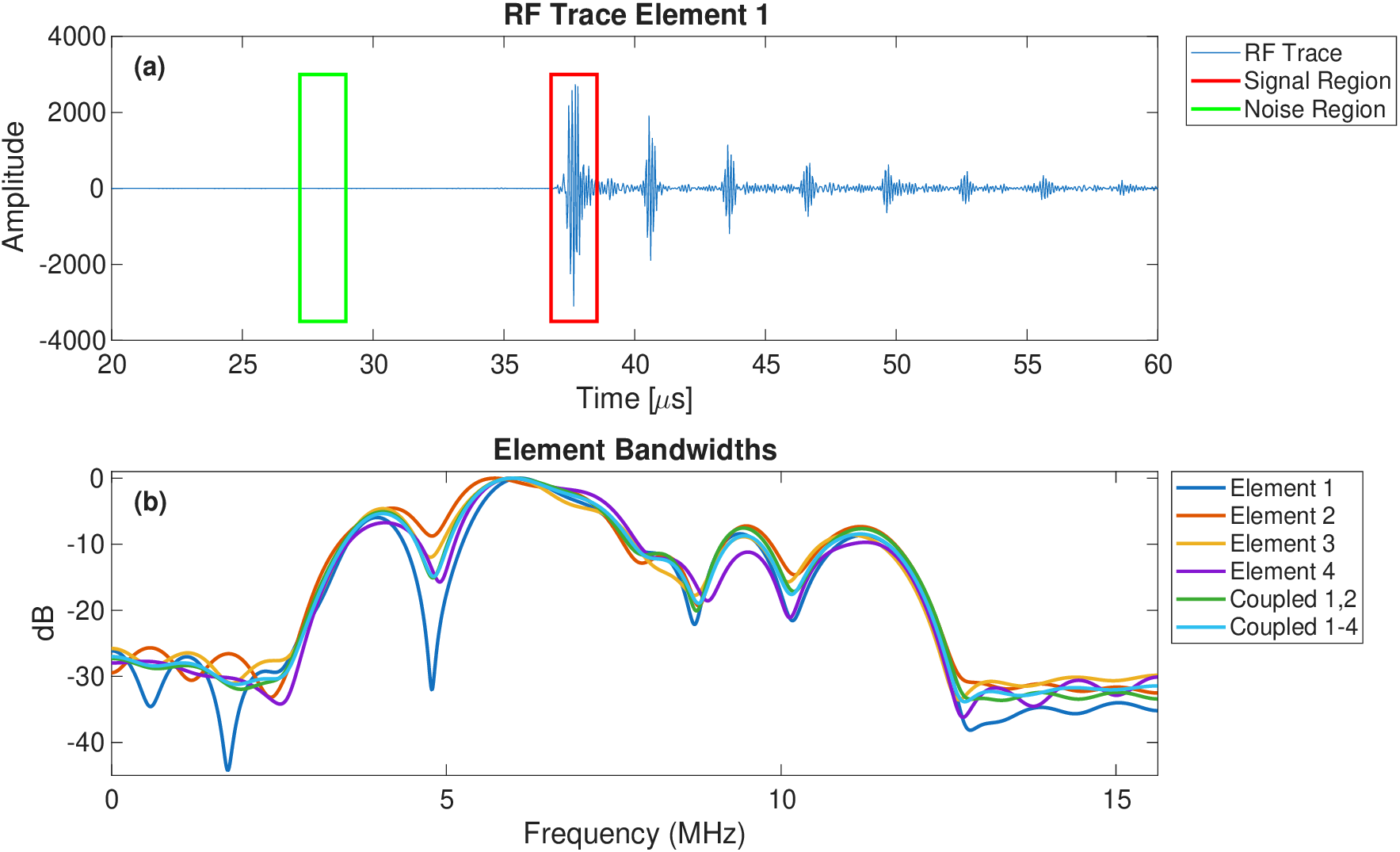}
    \caption{(a) Example RF trace displaying the signal and noise regions used for SNR estimates. The signal region was also used for bandwidth estimates. (b) Fourier transforms of the signal region of RF traces from the L14-5/60.}
    \label{fig6:bandwidth_traces}
\end{figure}

Next, we examined the bandwidths for coupled elements. The bandwidth of an element depends on the acoustic damping of the backing layer and the impedance matching between the backing, piezo-electric, and matching layers in the probe \cite{rathod_review_2020}. To examine how element coupling affects bandwidth, we estimated the bandwidths of the RF traces from the calibration dataset by taking their Fourier transform. An example RF trace is given in Figure \ref{fig6:bandwidth_traces}a. With element coupling, RF traces from adjacent elements are averaged together. When the traces are averaged, the Fourier transforms of the traces are averaged because the Fourier transform is linear. We observed slightly different bandwidths and center frequencies in the individual received traces, which would then average together during coupling. A few example Fourier transform magnitudes are displayed in Figure \ref{fig6:bandwidth_traces}b. The resulting bandwidths for individual elements 1, 2, 3, and 4, as well as coupled elements 1-2 and 1-4, are given in Table \ref{tab1:bandwidths}. The averaging of bandwidths is not a significant change as long as all individual elements in the probe have approximately the same center frequency and bandwidth. Ideally this is the case, based on our observations, and the fact that all elements are made of the same piezo-electric material and have the same backing and matching layers. A single large element made with the same layers will also have a comparable bandwidth. 

Finally, we examined the relative SNR values of individual versus coupled elements. The SNR was estimated by taking the mean envelope value of the signal region in Figure \ref{fig6:bandwidth_traces}a over the standard deviation of the noise region illustrated in the same Figure. We observed a slight SNR increase in coupled elements over the average individual element, likely due to averaging out electronics noise in the RF traces when summed during coupling. For an individual large element, we might expect the SNR to be more comparable to that of individual elements (e.g. 66-70 dB, rather than 75 dB), because no such averaging would take place.

\begin{table}[t]
    \centering
    \begin{tabular}{|c|c|c|c|c|c|c|}
        \hline
        Element & 1 & 2 & 3 & 4 & C 1-2 & C 1-4 \\
        \hline
        Bandwidth (MHz) & 1.99 & 2.37 & 2.30 & 2.31 & 2.23 & 2.27 \\
        SNR (dB) & 74.4 & 70.9 & 66.7 & 67.4 & 74.1 & 75.7 \\
        \hline
    \end{tabular}
    \caption{Element bandwidths and SNR values. The bandwidths were estimated from the -6 dB points of the time-domain Fourier Transforms of received planar reflector data. The SNR values were estimated from the ratio of the mean envelope of a planar reflector signal over the standard deviation in a noise region. The final two columns ``C 1-2" and ``C 1-4" denote coupled elements.}
    \label{tab1:bandwidths}
\end{table}

\section{Methods}

\subsection{Experiment setup}

\subsubsection{Element Coupling}
For all experiments, an Ultrasonix L14-5/60 probe was used, connected to a Verasonics Vantage 128 system (Verasonics, Inc., Kirkland, WA, USA). Coupling factors of 1, 2, and 4 were tested to divide evenly the 128-element array. These coupling factors resulted in element widths of 2.5 wavelengths, 5 wavelengths, and 10 wavelengths respectively when operating the probe at its standard center frequency of 7.81 MHz. To see the effects of element coupling on resolution and contrast, we scanned a CIRS Model 539 ATS General purpose phantom (Computerized Imaging Reference Systems, Norfolk, VA, USA). Data was acquired of wire targets and anechoic regions using two different angle sets. One set was the optimal angle set given in Table \ref{tab2:angle_sets}. The other set of phantom scans used the same angle set for each coupling factor. That angle set was the optimal set for a coupling factor of four, i.e. the most restricted angle set (see the third row of Table \ref{tab2:angle_sets}). This way, we could observe effects on image quality resulting purely from the directivity/element width. Additionally, we scanned the abdomen of a New Zealand White Rabbit to examine the effects of coupling in vivo. Animal procedures were approved by the Institutional Animal Care and Use Committee (IACUC) at the University of Illinois at Urbana-Champaign. Rabbits were anesthetized using isoflurane, then the fur on the rabbit's abdomen was shaved for imaging.

\subsubsection{Increased aperture size}
\begin{figure}
    \centering
    \includegraphics[width=\linewidth]{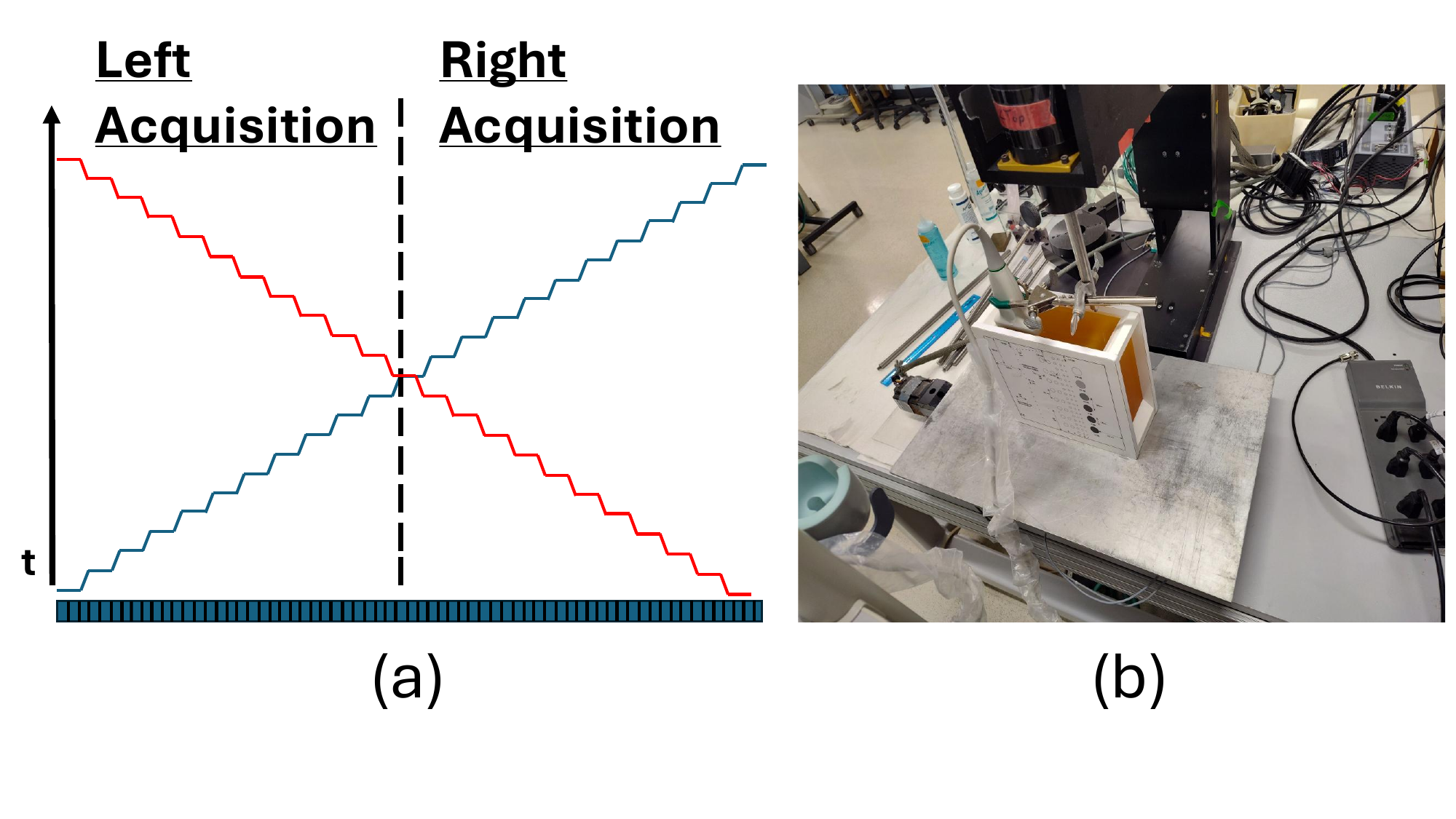}
    \caption{(a) A diagram of possible transmission profiles for either half of the virtual large-aperture acquisition with coupled elements. Blue represents a plane-wave steered in the positive \(\theta\) direction, while red is the negative \(\theta\) direction. (b) Photograph of the imaging setup for acquiring data from a virtual large aperture. The probe was held in place while the phantom was moved on the sliding table beneath it.}
    \label{fig8:daedal_photo}
\end{figure}

This experiment was performed to demonstrate how larger elements can lead to larger apertures with reduced element counts. The basic idea was to move the phantom to two positions under the probe, so the probe could transmit and collect data as if it were sections of an aperture of twice the size (see Figure \ref{fig8:daedal_photo}a). The L14-5/60 probe and the ATS phantom were placed on a Daedal positioning system (Parker Hannifin Corp., Cleveland, Ohio, USA), where the probe was fixed in place, while the phantom was on a sliding table allowing it to move laterally underneath the probe (see Figure \ref{fig8:daedal_photo}b). A few centimeters of degassed water were placed in the top of the ATS phantom to ensure good acoustic transmission into the phantom. The L14-5/60 probe is made up of \(N=128\) elements with a pitch \(p=472\) \(\mu\)m. Therefore, the distance the probe had to move (or equivalently, the phantom underneath it) to be aligned as a different section was \(Np = 60.416\) mm. Transmit delay profiles were designed for a virtual aperture of 120 mm with 128 elements. Then, corresponding halves of the transmission profile were fired from the L14-5/60 with elements coupled by 2 for either position (see Figure \ref{fig8:daedal_photo}a). Received channel data was then concatenated to be beamformed as if it came from a single aperture. 

To demonstrate expected performance of the virtual aperture, we also simulated the full aperture using K-wave \cite{martin_simulating_2016,treeby_rapid_2018}. We recreated the ATS phantom in k-wave using a grid spacing of 50 \(\mu\)m. The reference sound speed was set to 1450 m/s to match the sound speed of the ATS phantom, and the density was set to 1000 kg/m\(^3\). Then, speckle scatterers were included by introducing uniformly distributed random sound speed perturbations of maximum 1\% the reference sound speed at every grid point. Wire targets were included as circles of radius 150 \(\mu\)m, with sound speed 2000 m/s and density 4000 kg/m\(^3\). Anechoic cysts were defined as circles of various radii without speckle scatterers, i.e. homogeneous regions with the reference sound speed and density. Finally, the attenuation of the medium was set to 0.5 dB/cm/MHz. We also matched all transducer specs as closely as possible to the virtual array subject to discretizations by the computational grid (e.g. 919 \(\mu\)m element width in the virtual array became 900 \(\mu\)m width in the simulation).

\subsubsection{Beamforming}
This study compares the performance of conventional DAS beamforming with NSI \cite{agarwal_improving_2019}, SCF \cite{camacho_phase_2009}, and MV \cite{synnevag_adaptive_2007} beamformers. The quality of each of these beamformers depends on tuning parameters which can be set by the user. For DAS beamforming, the array F-number was estimated from the -3 dB point of the element directivity \cite{perrot_so_2021}
\begin{equation}
    F = \frac{1}{2\tan(\alpha)}
\end{equation}
where \(\alpha\) is the the -3 dB point. 

Plane-wave angle sets were chosen for each coupling factor using the method proposed by \cite{montaldo_coherent_2009}. In this method, the plane-wave angle set is given by  
\begin{equation}
    \theta_{i} = \arcsin(i\lambda/L) \approx i\lambda/L
\end{equation}
for \(i = -(n_{t}-1)/2, ... , (n_{t}-1)/2\), where \(\lambda\) is the wavelength, \(L\) is the total aperture size, and \(n_{t}\) is the number of transmissions. The number of transmissions needed is given by 
\begin{equation}
    n_{t} = \frac{L}{\lambda F}.
\end{equation}
Resulting angle sets for each coupling factor on the L14-5/60 and the virtual large aperture are given in Table \ref{tab2:angle_sets}. In addition to these optimal angle sets, data was collected for coupling factors 1 and 2 using the same angle set as coupling by 4 (third row, Table \ref{tab2:angle_sets}). This way, we could observe the effects of directivity and F-number separately from the effects of reduced angular compounding.

\begin{table}[t] 
    \begin{tabular}{|c|c|c|c|c|}
        \hline
        Array & \(\theta_{max}\) (deg) & Step (deg) & \(n_{t}\) & F\\
        \hline
        L14-5/60 Coupling by 1  &  11 & 0.187 & 118 & 2.6\\
        L14-5/60 Coupling by 2  &  5.3 & 0.187 & 58 & 5.2\\
        L14-5/60 Coupling by 4  &  2.6 & 0.187 & 28 & 10.7\\
        Virtual Large Aperture &  5.4 & 0.0935 & 116 & 5.2 \\
        \hline
    \end{tabular}
    \caption{Table of angle sets for each coupling factor on the L14-5/60 and the virtual large aperture. The second column is the maximum steering angle. The third column is the angular step size. The fourth column is the number of plane-wave angles. The fifth column is the array F-number.}
    \label{tab2:angle_sets}
\end{table}

For NSI, the image quality is controlled by a DC offset, usually in the range 0.1-1 \cite{agarwal_improving_2019}. Lower DC offsets will lead to narrower main lobes, lower side lobes, reduced grating lobes, but also reduced speckle signal \cite{gardner_grating_2024}. For our study, we empirically chose a DC offset of 0.5 for all NSI images to balance resolution increase with maintaining speckle signal.

For the SCF beamformer, the amount of suppression is controlled by an exponent applied to the estimated coherence factor, where larger exponents increase suppression of incoherent signals \cite{camacho_phase_2009}. For our study, we empirically chose an exponent of 1, again to balance speckle suppression with resolution increase. 

For MV, the weight calculation can be controlled in two ways: 1) the choice of sub-array length for spatial smoothing when estimating the covariance matrix, and 2) a diagonal loading factor applied to the estimated covariance matrix \cite{synnevag_adaptive_2007}. Longer sub-arrays and smaller diagonal loading factors both increase side-lobe suppression for MV. In this study, we simply use recommended values from the original paper, namely a sub-array length of \(L = N/2\), where \(N\) is the receive sub-aperture size determined by the F-number. Then, the diagonal loading factor is set to \(\epsilon = \frac{1}{10L} \cdot trace\{R\}\), where \(R\) is the covariance matrix estimate.

\subsection{Quality metrics}
Image quality was evaluated using several metrics. The most important metric for our study is resolution, which we analyzed in multiple ways. The first was the full-width at half-max (FWHM), which was estimated from the width between the -6 dB points in lateral profiles of wire targets. The FWHM gave an estimate of resolution based on what a user would observe in a B-mode image. In addition to the post-envelope FWHM, we also estimated the -6 dB beamwidth of the RF beamformed magnitude before envelope detection. Because NSI usually depends on null subtraction after the envelope, we applied the three required apodizations and performed the null subtraction on the beamformed RF magnitudes instead to include it in this comparison. This gave a closer estimate to true system resolution for non-linear beamformers that can change the dynamic range of the envelopes. Lastly, to examine whether spatial frequencies are truly recovered, we quantified the K-space by taking the 2D FFTs of the envelope magnitudes and examining the lateral frequencies available in the K-space as well. We also estimated the axial resolution based on the axial envelope FWHM of wire targets. 

Other quality metrics estimate the contrast and noise characteristics of images. The most simple metric of contrast is the contrast ratio (CR), often simply referred to as ``contrast." This is simply the dB difference of means between some region of interest (ROI), such as an anechoic target, and a nearby region of the speckle background. This metric essentially represents the visual contrast of an image, describing what a human might observe. It is measured as
\begin{equation}
    C = 20\log_{10}\left(\frac{\mu_{i}}{\mu_{o}}\right)
\end{equation}
where \(C\) is the contrast, and \(\mu_{i}\) and \(\mu_{o}\) are the mean envelope values inside and outside a ROI respectively. A more objective metric for contrast is the generalized contrast-to-noise ratio (gCNR), defined from the overlapping area of the histograms of two regions \cite{rodriguez-molares_generalized_2020}. The gCNR is defined as
\begin{equation} \label{eq:gcnr}
    gCNR = 1 - \int_{-\infty}^{\infty} \min_{x}\left\{p_{i}(x),p_{o}(x)\right\}dx
\end{equation}
where \(p_{i}(x)\) and \(p_{o}(x)\) are the histograms of the envelopes in some ROI and the background respectively. The gCNR is resilient against dynamic range changes and speckle variance, and thus gives a more objective metric for evaluating contrast improvement between different beamforming algorithms \cite{rodriguez-molares_generalized_2020}. Speckle statistics were evaluated using the speckle signal-to-noise ratio (sSNR). This ratio is defined as 
\begin{equation}
    sSNR = |\Bar{A}| / \sqrt{var(A)}
\end{equation}
where \(\Bar{A}\) is the mean of some speckle region, and \(var(A)\) is the variance in that region. Fully developed speckle will have an sSNR value of 1.91 \cite{wagner_statistics_1983}. In addition to sSNR, we also performed Kolmogorov-Smirnov (K-S) testing at near, mid, and far fields to determine whether speckle remained Rayleigh distributed. The ROIs used for K-S testing were first decimated by 10 pixels in either direction to ensure statistical independence of samples. Finally, we estimate the signal-to-noise ratio (SNR) to observe coupled element sensitivity. These estimates were made by taking the ratio of mean speckle signal over the standard deviation of an anechoic region, which should primarily contain noise.

\section{Results}
\subsection{Element Coupling}
\begin{figure}
    \centering
    \includegraphics[width=\linewidth]{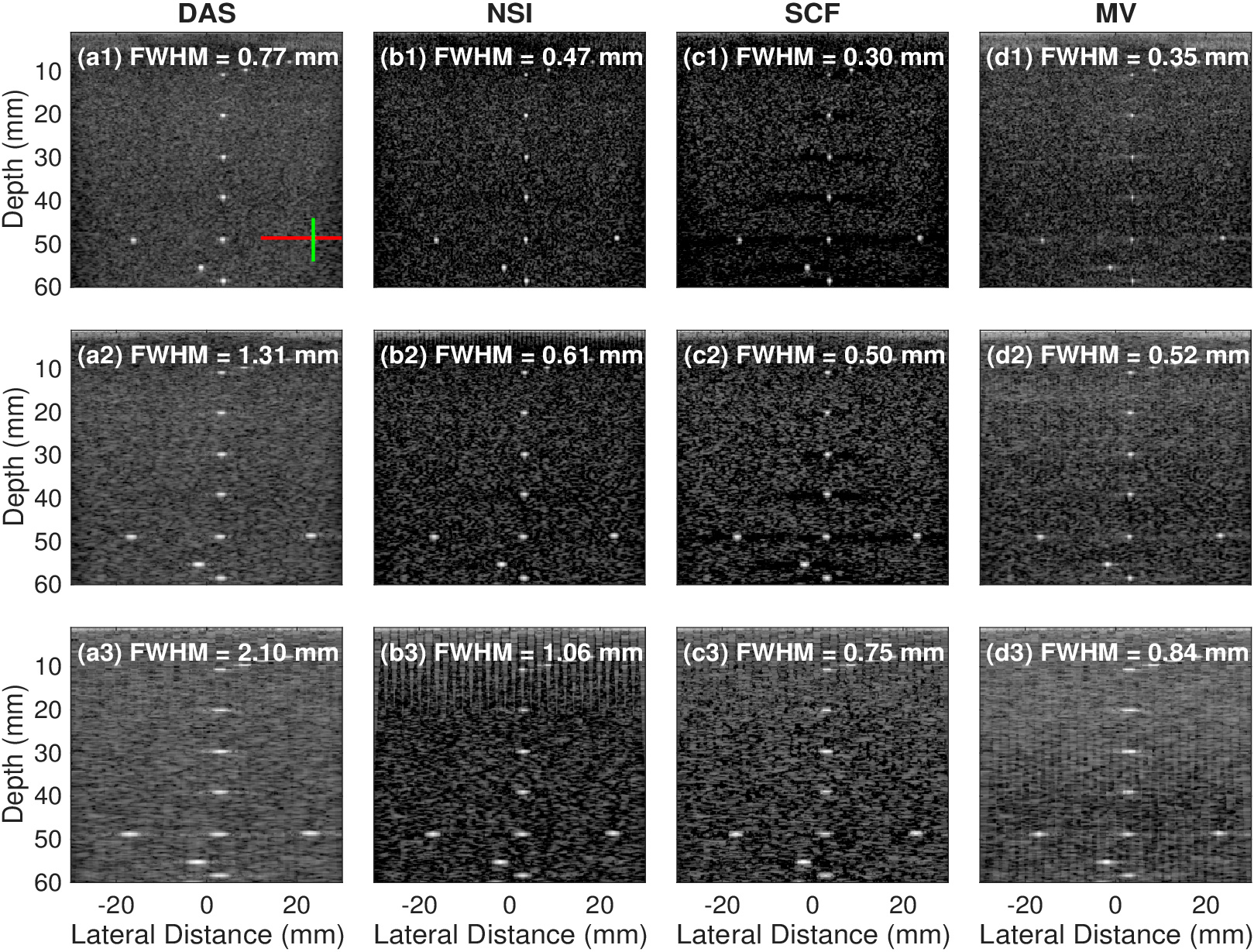}
    \caption{Wire phantom images from different beamformers and coupling factors. (a) DAS, (b) NSI, (c) SCF, (d) MV. (1) No coupling, (2) coupling by 2, (3) coupling by 4, which correspond to F-numbers of 2.6, 5.2, and 10.7 respectively. Each image is displayed with a dynamic range of 60 dB. The red and green lines in (a1) indicate the regions for lateral and axial profiles. The displayed FWHM values are lateral envelope FWHM estimates.}
    \label{fig9:wire_phantom_coupling}
\end{figure}

\begin{figure}
    \centering
    \includegraphics[width=\linewidth]{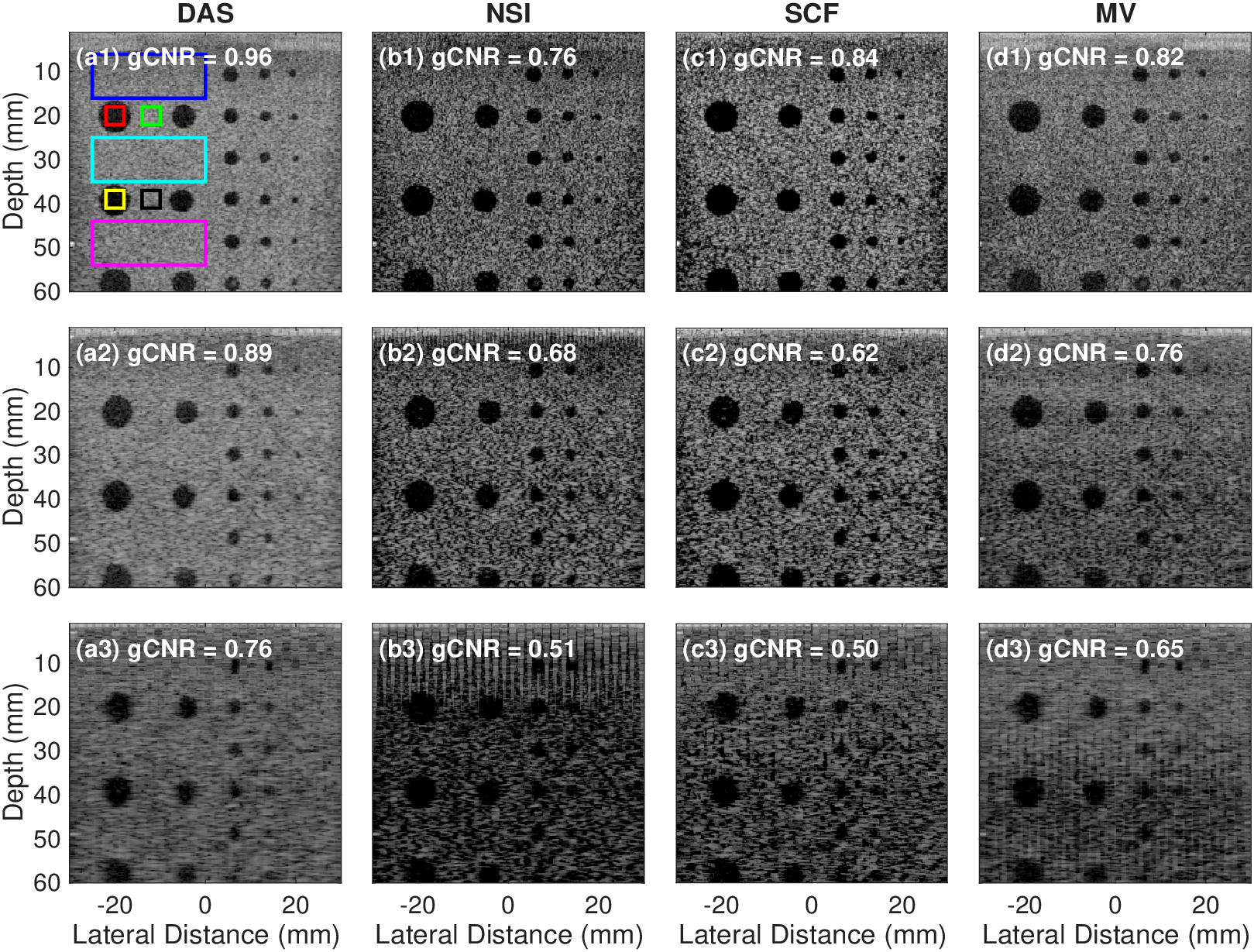}
    \caption{Cyst phantom images from different beamformers and coupling factors. (a) DAS, (b) NSI, (c) SCF, (d) MV. (1) No coupling, (2) coupling by 2, (3) coupling by 4, which correspond to F-numbers of 2.6, 5.2, and 10.7 respectively. Each image is displayed with a dynamic range of 60 dB. The ROIs for quality metrics are indicated in (a1). The red and green boxes indicate the regions for contrast ratio and gCNR. The blue, cyan, and magenta boxes were used for K-S testing, while the green box was used for sSNR. The yellow and black boxes were used for SNR.}
    \label{fig10:cyst_phantom_coupling}
\end{figure}

\begin{figure}
    \centering
    \includegraphics[width=\linewidth]{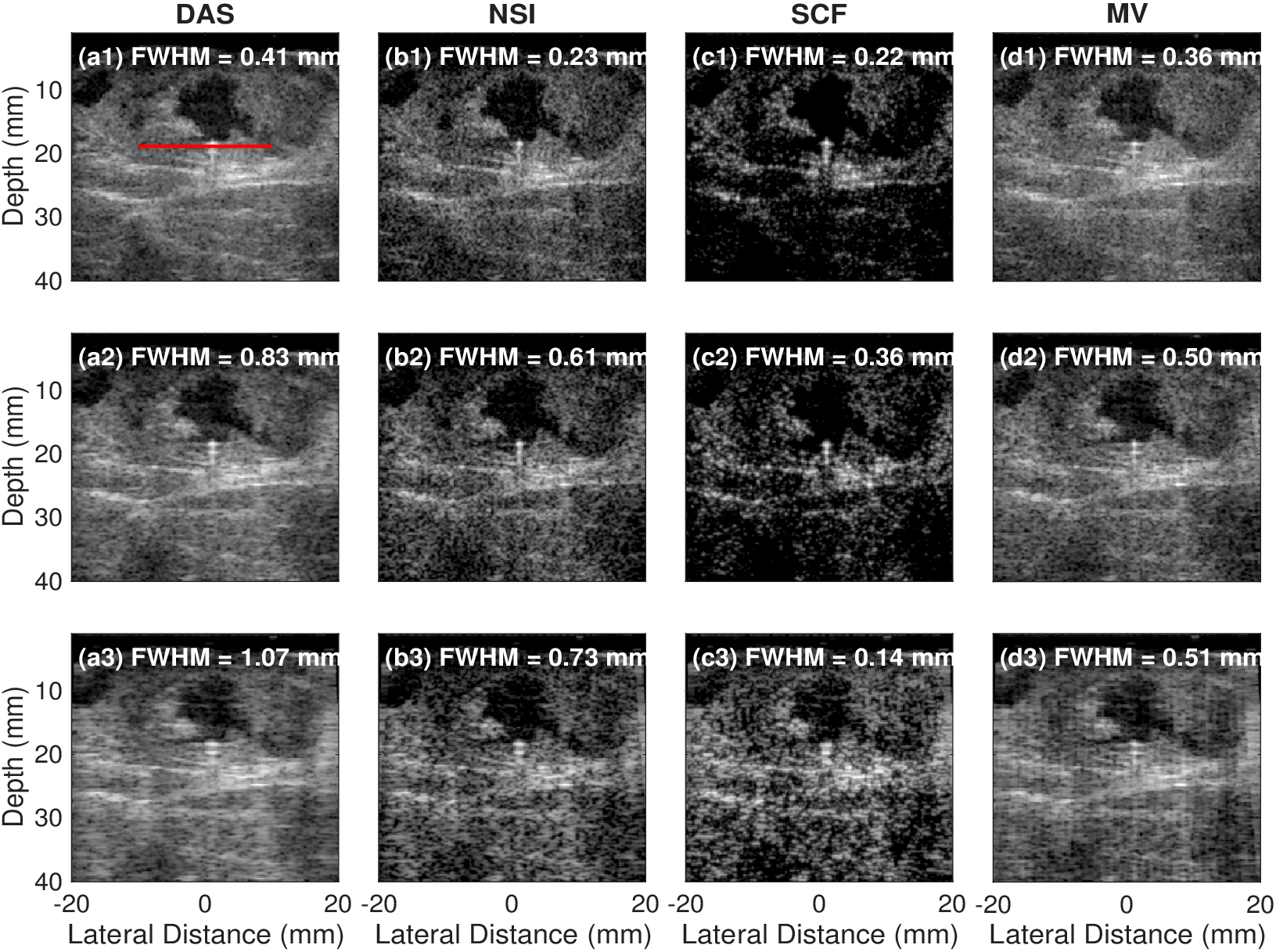}
    \caption{In vivo rabbit tumor images from different beamformers and coupling factors. (a) DAS, (b) NSI, (c) SCF, (d) MV. (1) No coupling, (2) coupling by 2, (3) coupling by 4, which correspond to F-numbers of 2.6, 5.2, and 10.7 respectively. Each image is displayed with a dynamic range of 60 dB. The ROI for the annotated FWHM values is indicated in (a1). The displayed FWHM values are lateral envelope FWHM estimates.}
    \label{fig11:in_vivo_figure}
\end{figure}

\begin{figure}
    \centering
    \includegraphics[width=\linewidth]{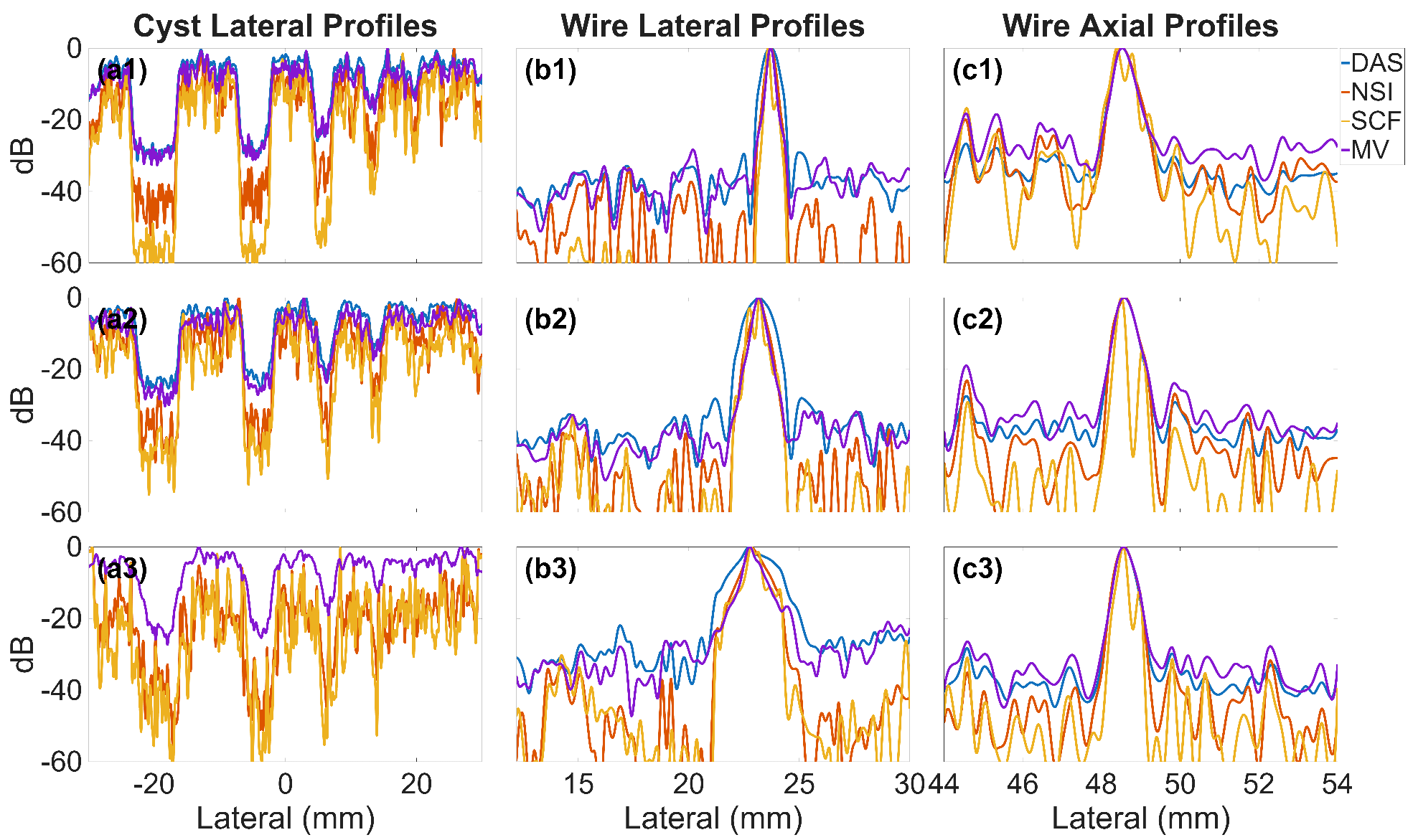}
    \caption{(a) Lateral profiles of anechoic regions. (b) Lateral profiles of wire targets. (c) Axial profiles of wire targets. (1) no coupling, (2) coupling by 2, and (3) coupling by 4. In each graph, the blue curve is DAS, the orange curve is NSI, the yellow curve is SCF, and the purple curve is MV.}
    \label{fig12:coupling_lateral_profiles}
\end{figure}

\begin{figure}
    \centering
    \includegraphics[width=\linewidth]{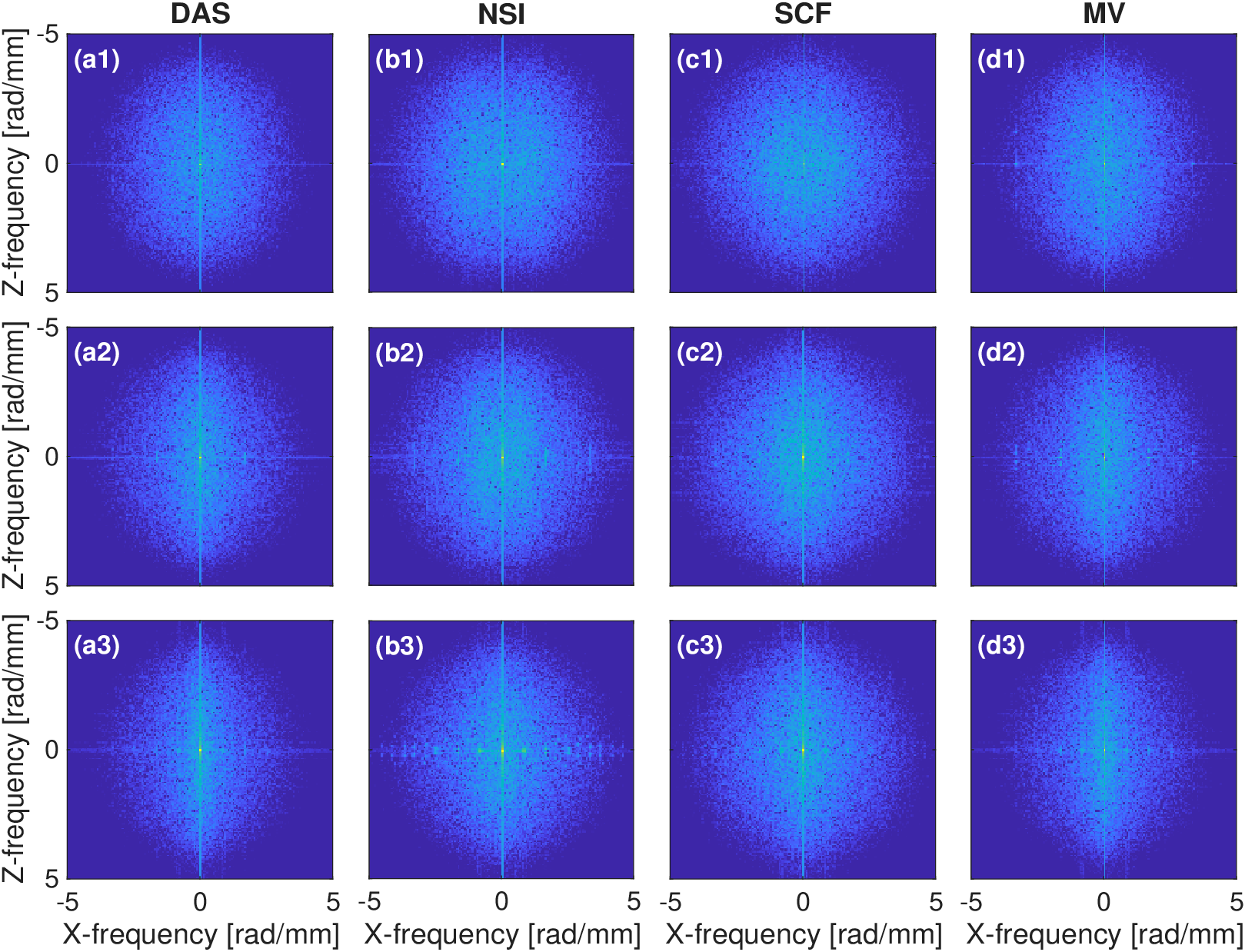}
    \caption{K-space support diagrams of the B-mode images in Figure \ref{fig9:wire_phantom_coupling}. (1) No coupling, (2) coupling by 2, and (3) coupling by 4. (a) DAS, (b) NSI, (c) SCF, (d) MV beamforming. Each diagram shows a dynamic range of 80 dB.}
    \label{fig13:kspace_diagrams}
\end{figure}

\begin{figure}
    \centering
    \includegraphics[width=\linewidth]{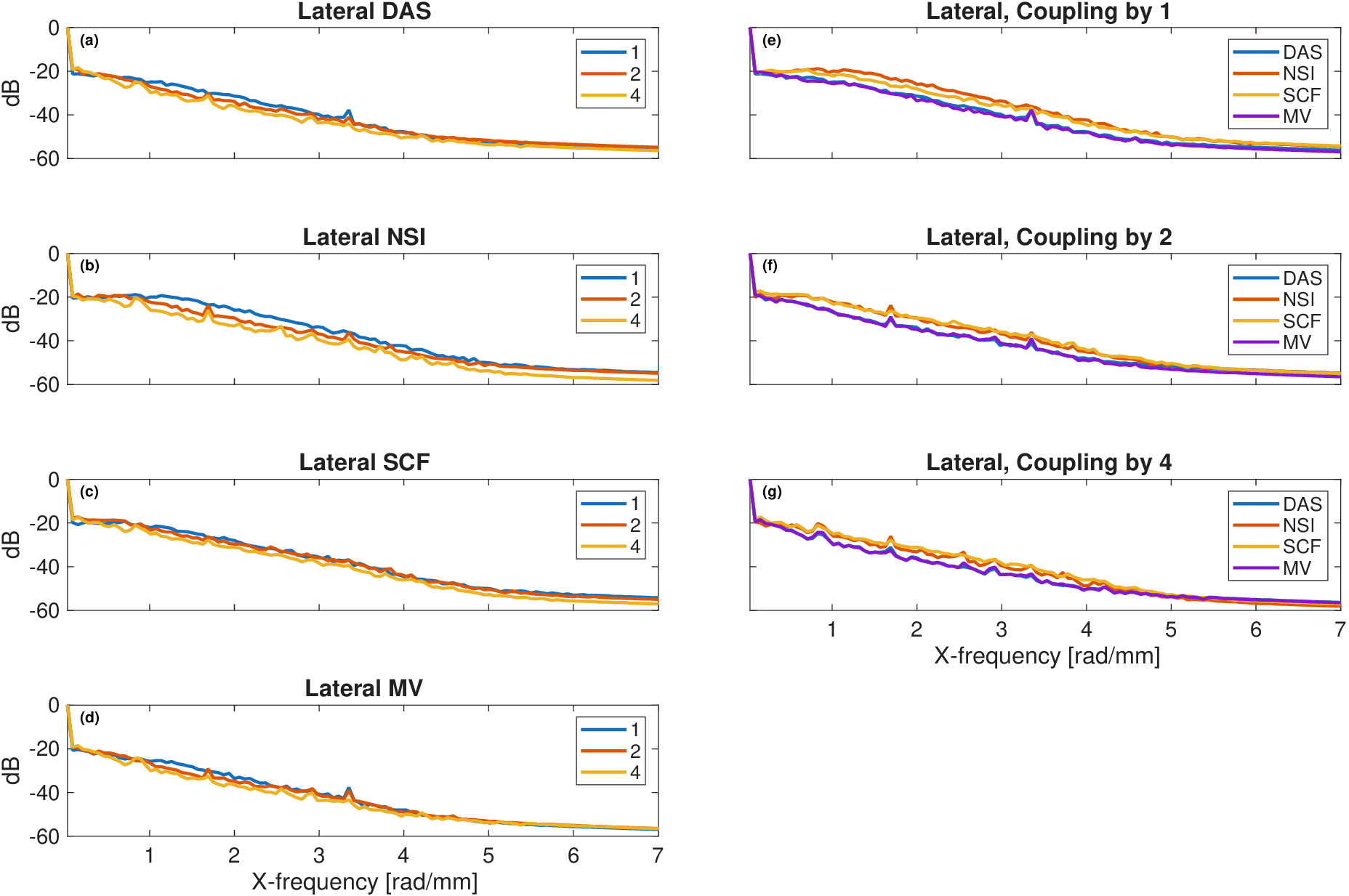}
    \caption{Lateral spatial frequency support, estimated by axially averaging the K-space magnitudes in Figure \ref{fig13:kspace_diagrams}. The left column displays each coupling factor with separate plots for beamformers, while the right column displays each beamformer with separate plots for coupling factors to facilitate comparison in both ways.}
    \label{fig14:kspace_lateral}
\end{figure}

The effects of coupling on anechoic targets, wire targets, and in vivo rabbit tumor images can be observed in Figures \ref{fig9:wire_phantom_coupling}, \ref{fig10:cyst_phantom_coupling}, and \ref{fig11:in_vivo_figure}, respectively. Lateral profiles for cysts and wires, and axial profiles for wires, are displayed in Figure \ref{fig12:coupling_lateral_profiles}. As general trends, reduced angular compounding resulted in a small loss to resolution while increased coupling resulted in much greater losses to resolution. For DAS beamforming, lateral envelope FWHM estimates for the wire target at the bottom right increased from 0.78 mm to 0.96 mm between the optimal and reduced angle sets (Fig. \ref{tab3:coupling_quality_metrics}a). The lateral FWHM estimate further increased to 1.8 mm and 2.4 mm with coupling factors of two and four. With alternative beamformers, lateral resolution was adjusted back to 1.2 mm with NSI, 0.76 mm with SCF, and 1.1 mm with MV for coupling by 4 on the reduced angle sets. The SCF beamformer had lower lateral FWHM estimates, but that was only due to distortion in the lateral profile of the wire, which can be viewed in Figure \ref{fig12:coupling_lateral_profiles}. In addition, the NSI beamformer resulted in the lowest lateral RF magnitude FWHM estimates, with all coupling factors being less than 0.5 mm. For coupling of 2, all advanced beamformers exceeded the lateral resolution of uncoupled DAS for both angle sets. Examining lateral resolution in the K-space diagrams gives further insights. In Figure \ref{fig13:kspace_diagrams}, the lateral frequency support is reduced for higher coupling factors, but is widened again by NSI and SCF beamforming. In Figure \ref{fig14:kspace_lateral}, all beamformers experienced reduced lateral frequency support with higher coupling factors. Also, the MV beamformer appears to have the same lateral frequency support as DAS for all coupling factors, while NSI and SCF both increase available lateral frequencies.

\begin{table*}[ht]
	\centering
	\begin{tabular}{|c|c|c|c|c|c|c|c|}
		\hline
		Beamformer & FWHM & RF FWHM & Ax FWHM & Contrast & gCNR & sSNR & SNR\\
		\hline
		DAS C1 & [0.96, 0.77] & [1.12, 0.84] & [0.48, 0.50] & [-23.60, -19.94] & [0.96, 0.96] & [1.21, 1.18] & 8.45\\
		NSI C1 & [0.47, 0.47] & [0.19, 0.16] & [0.49, 0.53] & [-25.25, -23.82] & [0.69, 0.76] & [0.61, 0.75] & 3.62\\
		SCF C1 & [0.41, 0.30] & [0.34, 0.26] & [0.15, 0.33] & [-40.82, -36.61] & [0.77, 0.84] & [0.40, 0.64] & 12.64\\
		MV C1 & [0.36, 0.35] & [0.49, 0.38] & [0.43, 0.44] & [-22.14, -18.83] & [0.92, 0.82] & [1.03, 0.92] & 5.99\\
		\hline
		DAS C2 & [1.77, 1.31] & [1.65, 1.47] & [0.49, 0.48] & [-24.18, -16.56] & [0.96, 0.89] & [1.28, 1.23] & 8.36\\
		NSI C2 & [0.79, 0.61] & [0.23, 0.13] & [0.50, 0.49] & [-29.24, -20.45] & [0.77, 0.68] & [0.69, 0.76] & 4.66\\
		SCF C2 & [0.44, 0.50] & [0.61, 0.52] & [0.24, 0.19] & [-34.00, -22.52] & [0.80, 0.62] & [0.52, 0.67] & 10.89\\
		MV C2 & [0.46, 0.52] & [0.64, 0.60] & [0.51, 0.48] & [-24.88, -16.25] & [0.94, 0.76] & [1.08, 0.98] & 6.40\\
		\hline
		DAS C4 & 2.41 & 1.86 & 0.49 & -21.14 & 0.94 & 1.36 & 6.08\\
		NSI C4 & 1.14 & 0.35 & 0.49 & -21.99 & 0.70 & 0.69 & 3.49\\
		SCF C4 & 0.70 & 1.09 & 0.18 & -21.85 & 0.74 & 0.87 & 3.01\\
		MV C4 & 1.07 & 1.43 & 0.49 & -21.14 & 0.94 & 1.36 & 4.76\\
		\hline
	\end{tabular}
	\caption{Quality metrics, given as [reduced angles, optimal angles]. All FWHM estimates are in mm. Contrast and SNR are in dB. C1, C2, and C4 indicate coupling factors. In the C4 rows, the reduced and optimal angle sets were the same.}
	\label{tab3:coupling_quality_metrics}
\end{table*}

As for contrast, element coupling also reduced the contrast ratio from -24 dB without coupling to -18 dB coupling by 4, while gCNR moved from 0.98 without coupling to 0.90 coupling by 4. In our datasets, using more angles actually decreased the contrast, going from -19.9 dB with the optimal angles down to -23.6 dB with the reduced angle set. The MV beamformer did not improve either contrast metric for any coupling factor. Meanwhile, the NSI and SCF beamformers improved the contrast ratio, but lowered the gCNR. The best contrast ratio on coupled elements was achieved by the SCF beamformer for coupling by 2, with a value of -25 dB. Yet SCF also resulted in the lowest gCNR for coupling by 2, with a value of 0.69. In general, we observed that while alternative beamformers improved the contrast ratio for every coupling factor, they all also lowered the gCNR. 

Element coupling slightly increased the sSNR values, going from 1.18 without coupling to 1.33 with coupling using DAS. The sSNR estimates were lower with the alternative beamformers, the lowest being 0.72 coming from NSI at a coupling factor of 4. The highest sSNR values among the alternative beamformers came from MV. As a complement to the sSNR values, the K-S test results are given in Table \ref{tab4:kstest}. From these tests, the DAS beamformer produced Rayleigh-distributed speckle at all imaging regions. Meanwhile, NSI and SCF never produced Rayleigh-distributed speckle, and MV only produced Rayleigh-distributed speckle for higher coupling factors. The K-S test results appear to line up reasonably with the sSNR values, where the highest sSNR values pass the K-S test as Rayleigh-distributed speckle, while lower sSNR values (e.g. less than 1) are not Rayleigh distributed.

\begin{table}[t]
    \centering
    \begin{tabular}{|c|c|c|c|c|}
        \hline
        Coupling Factor &  DAS & NSI & SCF & MV\\
        \hline
        1 & [1,1,1]& [0,0,0]& [0,0,0]& [0,0,0]\\
        2 & [1,1,1]& [0,0,0]& [0,0,0]& [1,1,0]\\
        4 & [1,1,1]& [0,0,0]& [0,0,0]& [1,1,1]\\
        \hline
    \end{tabular}
    \caption{K-S test results for all coupling factors. A 1 indicates pass (Rayleigh distributed speckle) while a 0 indicates fail (non-Rayleigh speckle). Entries are given as [near, mid, far] field.}
    \label{tab4:kstest}
\end{table}

There was an increase in the size of a ``dead zone" near the transducer face for higher coupling factors. This was manifest as the spike artifacts at the top of the images in Figures \ref{fig9:wire_phantom_coupling} and \ref{fig10:cyst_phantom_coupling}. The exact depth of the dead zone, which we defined as the depth at which at least two element contribute, is given in Table \ref{tab5:dead_zone_size}. This depth was calculated with simple geometry, using the element widths and the acceptance angle to find the depth where adjacent element directivities would overlap. 

Analyzing the axial resolution, the coupling factors do not have much effect. Many of the axial FWHM values in Figure \ref{tab3:coupling_quality_metrics}e overlap with each other (e.g. DAS, NSI, and MV all overlap for coupling by 4). SCF has apparently better axial resolution, but only because SCF distorts the signal by splitting the wire into two apparent targets (see Figures \ref{fig12:coupling_lateral_profiles}c1-\ref{fig12:coupling_lateral_profiles}c3), which is not accurate.

\subsection{Increased aperture size}
\begin{figure*}
    \centering
    \includegraphics[width=\linewidth]{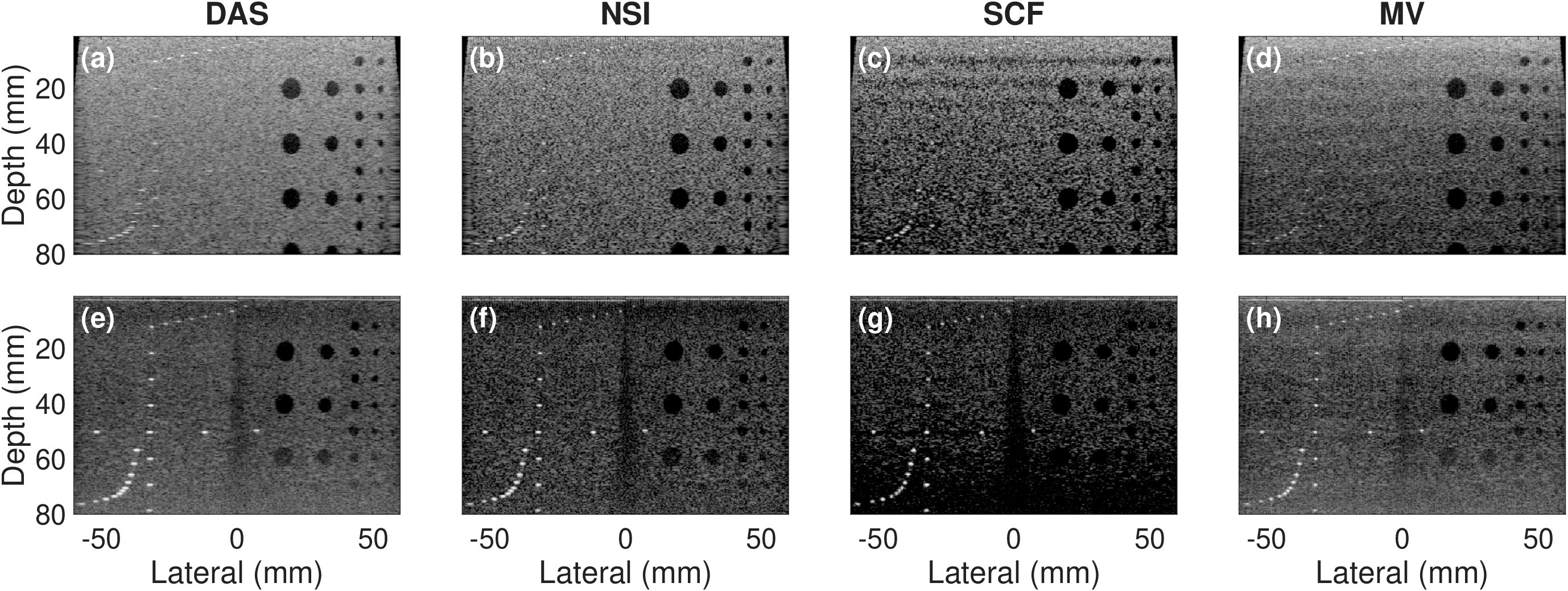}
    \caption{B-mode images of a virtual large aperture made up of 128 elements with an element width of 5 wavelengths, corresponding to an F-number of 5.2. (a)-(d) Simulated large aperture from K-wave. (e)-(h) Acquired large aperture with the positioning system.}
    \label{fig15:virtual_large_aperture_image}
\end{figure*}

\begin{figure}
    \centering
    \includegraphics[width=\linewidth]{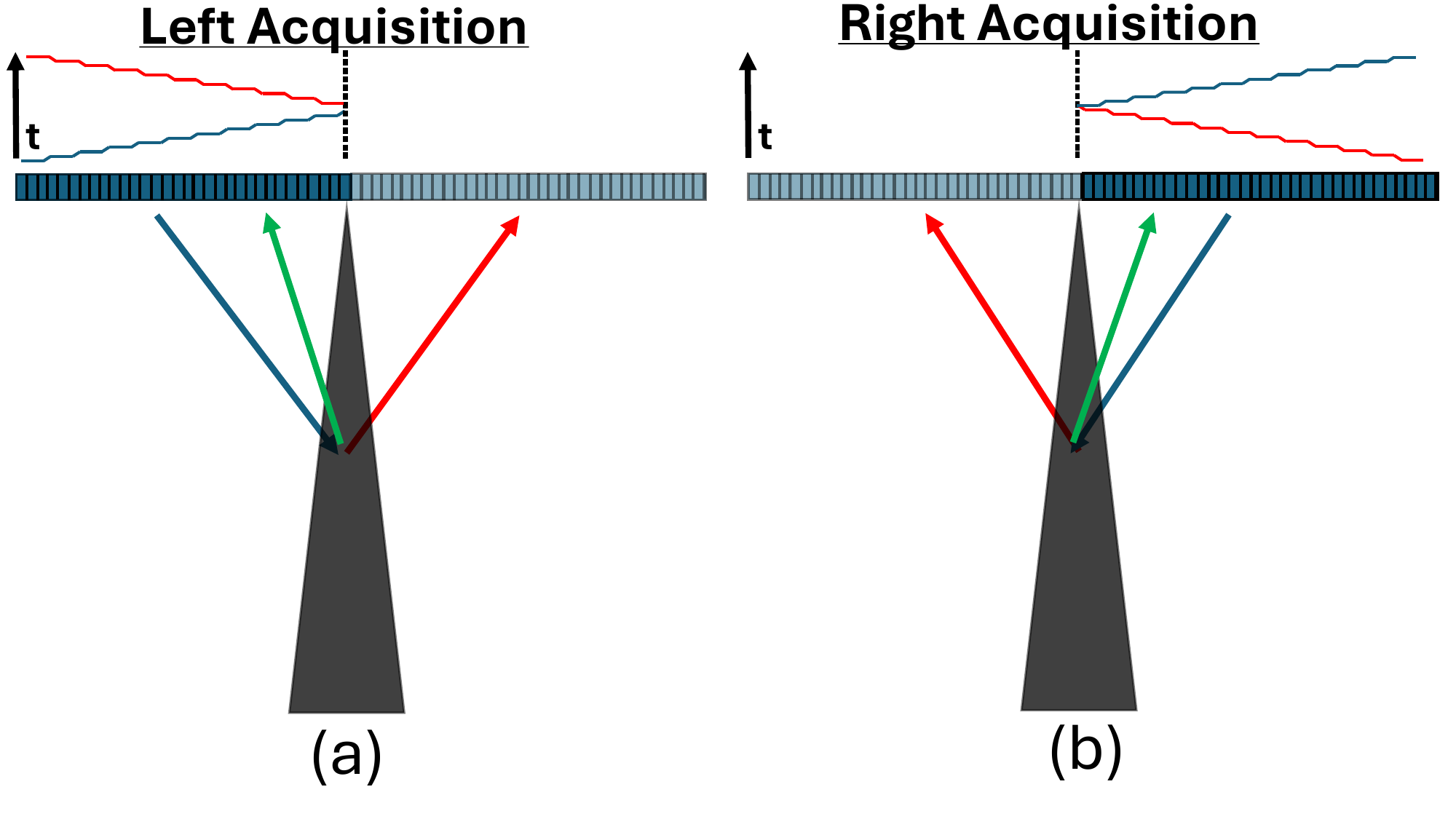}
    \caption{Diagrams of missing transmit/receive pairs for acquisitions on either position of the virtual large aperture. The blue arrows represent transmit paths. The green arrows represent successfully received echos. The red arrows are echos that cannot be received by the virtual array due to probe position.}
    \label{fig16:virtual_aperture_digram}
\end{figure}

The resulting B-mode images of the virtual large aperture are displayed in Figure \ref{fig15:virtual_large_aperture_image}. Figure \ref{fig15:virtual_large_aperture_image}a-d display the K-wave simulation result, while Figure \ref{fig15:virtual_large_aperture_image}e-h came from the acquired data using the positioning system. Each of these images has a 120 mm FOV, double that of the L14-5/60, but the same number of elements as the 14-5/60. Many of the same quality observations comparing the beamformers can be made here as were made in the previous section. The main new observation is the dark cone in the center of the acquired images (Figure \ref{fig15:virtual_large_aperture_image}e-h). This is an artifact of the acquisition setup, where data from either half of the virtual large aperture was collected separately. There was no way for the left half to transmit while the right half received or vice-versa, but pixels in the center rely on data from both halves during beamforming. That data was missing due to the limitations in our setup, creating this artifact (see Figure \ref{fig16:virtual_aperture_digram}). However, this artifact would not be present in a built large aperture because it would not have such acquisition constraints. This can be observed in Figure \ref{fig15:virtual_large_aperture_image}a-d, where the simulated large aperture did not produce this artifact.

\section{Discussion}
In our approach to increasing FOV, the most important consideration is the size of the elements and the effects that element width has on image quality. As predicted by the theoretical beam patterns, resolution decreased greatly for larger elements (i.e. higher coupling factors) due to the narrowed directivity and increased F-number. In comparing the influence of reduced compounding versus increased F-number, we found that increased F-number resulted in a much greater loss to resolution than a limited steering range. The image contrast was fairly consistent across coupling factors, especially with the gCNR metric. The speckle was also fairly consistent, with all coupling factors producing Rayleigh-distributed speckle on DAS. Higher coupling also slightly increased the sSNR metric, likely due to a larger resolution cell containing more scatterers, thus producing more fully developed speckle. Higher coupling factors had no effect on axial resolution. This is because axial resolution does not depend on the directivity/F-number, rather it depends on excitation pulse length (which is held constant throughout the study) and transducer bandwidth (which, from the discussion in Section II.B, does not significantly change with coupling). Finally, the SNR of speckle signal versus an anechoic region was actually reduced a little by higher coupling factors. With larger elements, one might expect that SNR would be higher because single large elements have higher sensitivity (i.e. they transmit/receive more power) than single small elements. However, the reduction to SNR is likely another side effect of a narrowed directivity. On receive, a smaller proportion of the total aperture can effectively contribute to beamforming the pixel, meaning fewer signals are averaged to calculate the pixel value, allowing more noise to come through. 

In addition to resolution and contrast effects, higher coupling factors also created a ``dead zone" near the transducer face. The dead zone was manifest as spikes that appear near the top of the B-mode images in Figures 7, 8, and 9 for higher coupling factors. For these pixels, only one or two elements can contribute due to narrow directivities. With only one element, no beamforming can be done. In this case, the images simply display the envelope of the time-delayed sample value from that element. These are the downward facing spikes. With two elements, now the beamformers can have an effect, often significantly changing the value compared to adjacent pixels which only came from one element. These are the upward facing spikes. To quantify the dead zone size, a simple trigonometric calculation can be set up using the distance between two elements and the acceptance angle of the elements to find the depth where at least two elements contribute to the pixels. These depths are enumerated in Table \ref{tab5:dead_zone_size} for each coupling factor. For many abdominal applications, such as liver or kidney imaging, the dead zone is not a major concern because those tasks often require a great imaging depth, especially with obese patients. In these cases, the region of interest should be well past the dead zone. For other cases, such as tumor imaging, if the tumor is superficial, the dead zone can be a serious limitation. However, this limitation could be mitigated with the use of standoffs or transparent gel pads to create distance between the transducer and a superficial region of interest \cite{tsui_flexible_2012,corvino_utility_2020}.

\begin{table}[t]
    \centering
    \begin{tabular}{|c|c|}
        \hline
        Coupling Factor & Dead Zone Depth \\
        \hline
        1 & 1.3 mm\\
        2 & 5.29 mm\\
        4 & 21.43 mm\\
        \hline
    \end{tabular}
    \caption{Depth of the dead zone for each coupling factor, defined by the depth past which at least two elements contribute.}
    \label{tab5:dead_zone_size}
\end{table} 

Using a coupling factor of two, we were able to produce B-mode images with high resolution from a virtual large aperture of 120 mm width and 128 elements, demonstrating how larger apertures could be built using larger elements without increase to element count. Our goal is to take this approach to 2D matrix arrays, creating larger apertures using large square elements, and the results demonstrate for a 1D case the effectiveness of the approach. Moving towards 2D will have many of the same considerations, and a few added complexities that are worth discussing. First, the directivities of square elements are a simple, orthogonal extension of the directivities of line elements. A square element can be thought of as two orthogonal rectangle functions, so its directivity can be modeled as two orthogonal sinc functions. The directivity considerations will largely be the same in 2D as they were for the 1D case, where larger elements increase directivity, raising the F-number and reducing resolution in both azimuth and elevation directions. There is a potential for grating lobes in both azimuth and elevation, as well as along element diagonals because matrix arrays are also periodic along diagonals. Another consideration is that matrix arrays have much higher element counts overall, often requiring multiplexers or micro-beamformers as mentioned in the introduction. Our designs will aim to bring down element counts enough to avoid added hardware complexity. Aside from array design, directional implementations of each beamformer will be required (e.g. directional zero-mean apodizations for NSI). Lastly, spatial compounding can also become more complex, requiring plane-wave angles to be steered in both azimuth and elevation directions. An optimal angle set derived from Montaldo's method \cite{montaldo_coherent_2009} may require a squared number of angles (i.e. every azimuth/elevation combination), far too many to maintain a reasonable frame rate, even with large elements and a limited steering range. Angle decimation will likely be required, for which case, compressed sensing methods such as tensor completion \cite{afrakhteh_coherent_2021} or radial basis functions \cite{afrakhteh_two-dimensional_2023} could be used to interpolate decimated angle sets and maintain high quality. 

The other major consideration in our approach is the beamforming method used to maintain resolution and understanding the trade-offs of the chosen method. The major advantages of NSI were that it had the best RF FWHM values and it increased spatial frequency support in the K-space diagrams over DAS. NSI is also a relatively inexpensive beamformer in terms of computation \cite{agarwal_improving_2019}. However, its major disadvantages were in speckle quality and contrast. NSI did not produce Rayleigh distributed speckle in any region for any coupling factor, and produced much lower sSNR values than DAS or MV. NSI also produced the lowest gCNR metrics. On the other hand, the MV beamformer produced the best envelope FWHM values, best gCNR and sSNR behind DAS, and maintained Rayleigh distributed speckle for coupling by two and four. The major disadvantage with MV is that it is the most computationally expensive of the beamformers we tried because its calculation involves a matrix inversion \cite{synnevag_adaptive_2007}. The SCF beamformer also produced low envelope FWHM, and like NSI maintained a higher spatial frequency support than DAS or MV. It also produced the best contrast ratio estimates. However, SCF resulted in distortion of wire targets and never produced Rayleigh-distributed speckle according to K-S tests. 

As a final consideration, with these beamformers, it is important to point out that parameter tuning affects the quality of their results significantly. Our particular values of tuning parameters were 1) a DC offset of 0.5 for NSI, 2) an exponent of 1 for SCF, and 3) sub-array size \(L=N/2\) and diagonal loading factor \(\epsilon = \frac{1}{10L} \cdot trace\{R\}\) for MV. The parameters for MV were chosen based on recommended values in the paper which introduced it for ultrasound B-mode imaging \cite{synnevag_adaptive_2007}. The parameters for NSI and SCF were chosen empirically to balance speckle suppression with resolution increase. While our results are limited to our chosen set of tuning parameters, the trade-offs of different tunings are known and already published in the literature \cite{agarwal_improving_2019,synnevag_adaptive_2007,camacho_phase_2009}. In particular, in tuning for a resolution increase, all three beamformers will also result in higher speckle variance, which can lead to underdeveloped speckle and reduce image contrast.

\section{Conclusion}
This study was aimed at increasing ultrasound FOV using larger elements and adaptive or non-linear beamforming. We have demonstrated through theoretical analysis how a minimum kerf is optimal to minimize grating lobes for an increased pitch. We also demonstrated that coupled elements make a close approximation of large elements in directivity, sensitivity, and bandwidth. Experiments were performed with coupled elements in phantoms and in vivo that demonstrated the resolution loss from an increased F-number with larger elements. However, our experiments also demonstrated how adaptive and non-linear beamformers, such as NSI, SCF, and MV could be used to not only regain resolution, but actually exceed the resolution of DAS beamforming on an uncoupled array. Using these alternative beamformers, high-resolution images were reconstructed with elements up to 5 wavelengths wide. Even elements up to 10 wavelengths wide could be used with the trade-off of requiring a standoff to bring superficial regions out of the dead zone. We have also directly demonstrated how larger elements can lead to larger apertures by collecting data from a virtual large aperture using a positioning system. This virtual aperture had a width of 120 mm with only 128 elements. We see potential application of our approach for increasing the FOV for abdominal imaging in 1D and for reducing the element counts of 2D matrix arrays.

\begin{acknowledgments}
This work was supported by the National Institutes of Health (NIH), grant numbers R01CA251939, R01CA273700, and R21EB024133
\end{acknowledgments}

\section*{Author Declarations}
The authors have no competing interests to declare. 

Animal procedures were approved by the Institutional Animal Care and Use Committee at the University of Illinois at Urbana-Champaign, protocol number 23062.

\section*{Data Availability}
Data will be made available upon request.

\bibliography{large_aperture_1d}

@article{bottenus_impact_2020,
	title = {The {Impact} of {Acoustic} {Clutter} on {Large} {Array} {Abdominal} {Imaging}},
	volume = {67},
	issn = {1525-8955},
	doi = {10.1109/TUFFC.2019.2952797},
	language = {eng},
	number = {4},
	journal = {IEEE transactions on ultrasonics, ferroelectrics, and frequency control},
	author = {Bottenus, Nick and Pinton, Gianmarco F. and Trahey, Gregg},
	month = apr,
	year = {2020},
	pages = {703--714},
}

@article{foiret_improving_2022,
	title = {Improving plane wave ultrasound imaging through real-time beamformation across multiple arrays},
	volume = {12},
	copyright = {2022 The Author(s)},
	issn = {2045-2322},
	doi = {10.1038/s41598-022-16961-2},
	language = {en},
	number = {1},
	urldate = {2023-11-28},
	journal = {Scientific Reports},
	author = {Foiret, Josquin and Cai, Xiran and Bendjador, Hanna and Park, Eun-Yeong and Kamaya, Aya and Ferrara, Katherine W.},
	month = aug,
	year = {2022},
	pages = {13386},
}

@article{kim_extended_2003,
	title = {Extended {Field}-of-{View} {Sonography}},
	volume = {22},
	copyright = {© 2016 by the American Institute of Ultrasound in Medicine},
	issn = {1550-9613},
	doi = {10.7863/jum.2003.22.4.385},
	language = {en},
	number = {4},
	urldate = {2023-11-28},
	journal = {Journal of Ultrasound in Medicine},
	author = {Kim, Se Hyung and Choi, Byung Ihn and Kim, Kyoung Won and Lee, Kyoung Ho and Han, Joon Koo},
	year = {2003},
	pages = {385--394},
}

@article{kang_wide_2020,
	title = {Wide {Field}-of-{View} {Ultrafast} {Curved} {Array} {Imaging} {Using} {Diverging} {Waves}},
	volume = {67},
	issn = {1558-2531},
	doi = {10.1109/TBME.2019.2942164},
	number = {6},
	urldate = {2023-11-28},
	journal = {IEEE Transactions on Biomedical Engineering},
	author = {Kang, Jinbum and Go, Dooyoung and Song, Ilseob and Yoo, Yangmo},
	month = jun,
	year = {2020},
	pages = {1638--1649},
}

@inproceedings{wang_wide_2022,
	title = {Wide {Field}-of-{View} {Plane} {Wave} {Ultrasound} {Imaging} based on {Linear} {Array} {Sub}-apertures and {Adaptive} {Weighting} {Technique}},
	doi = {10.1109/IUS54386.2022.9958863},
	urldate = {2023-11-28},
	booktitle = {2022 {IEEE} {International} {Ultrasonics} {Symposium} ({IUS})},
	author = {Wang, Yadan and Zheng, Chichao and Wang, Yuanguo and Liu, Mingzhou and Peng, Hu},
	month = oct,
	year = {2022},
	pages = {1--4},
}

@article{gavrilov_method_1997,
	title = {A method of reducing grating lobes associated with an ultrasound linear phased array intended for transrectal thermotherapy},
	volume = {44},
	issn = {1525-8955},
	doi = {10.1109/58.655626},
	number = {5},
	urldate = {2023-11-28},
	journal = {IEEE Transactions on Ultrasonics, Ferroelectrics, and Frequency Control},
	author = {Gavrilov, L.R. and Hand, J.W. and Abel, P. and Cain, C.A.},
	month = sep,
	year = {1997},
	pages = {1010--1017},
}

@inproceedings{camacho_grating-lobes_2009,
	title = {Grating-lobes reduction by application of {Phase} {Coherence} {Factors}},
	issn = {1948-5727},
	doi = {10.1109/ULTSYM.2009.5441770},
	urldate = {2023-11-28},
	booktitle = {2009 {IEEE} {International} {Ultrasonics} {Symposium}},
	author = {Camacho, J. and Parrilla, M. and Fritsch, C.},
	month = sep,
	year = {2009},
	pages = {341--344},
}

@article{agarwal_improving_2019,
	title = {Improving {Spatial} {Resolution} {Using} {Incoherent} {Subtraction} of {Receive} {Beams} {Having} {Different} {Apodizations}},
	volume = {66},
	issn = {1525-8955},
	doi = {10.1109/TUFFC.2018.2876285},
	number = {1},
	urldate = {2023-11-28},
	journal = {IEEE Transactions on Ultrasonics, Ferroelectrics, and Frequency Control},
	author = {Agarwal, Anil and Reeg, Jonathan and Podkowa, Anthony S. and Oelze, Michael L.},
	month = jan,
	year = {2019},
	pages = {5--17},
}

@article{kou_grating_2022,
	title = {Grating {Lobe} {Reduction} in {Plane}-{Wave} {Imaging} {With} {Angular} {Compounding} {Using} {Subtraction} of {Coherent} {Signals}},
	volume = {69},
	issn = {1525-8955},
	doi = {10.1109/TUFFC.2022.3217993},
	number = {12},
	urldate = {2023-11-28},
	journal = {IEEE Transactions on Ultrasonics, Ferroelectrics, and Frequency Control},
	author = {Kou, Zhengchang and Miller, Rita J. and Oelze, Michael L.},
	month = dec,
	year = {2022},
	pages = {3308--3316},
}

@misc{kou_high-resolution_2023,
	title = {High-resolution {Power} {Doppler} {Using} {Null} {Subtraction} {Imaging}},
	doi = {10.48550/arXiv.2301.03719},
	urldate = {2023-11-28},
	publisher = {arXiv},
	author = {Kou, Zhengchang and Lowerison, Matthew and You, Qi and Wang, Yike and Song, Pengfei and Oelze, Michael L.},
	month = jan,
	year = {2023},
}

@article{camacho_phase_2009,
	title = {Phase {Coherence} {Imaging}},
	volume = {56},
	issn = {1525-8955},
	doi = {10.1109/TUFFC.2009.1128},
	number = {5},
	urldate = {2023-11-28},
	journal = {IEEE Transactions on Ultrasonics, Ferroelectrics, and Frequency Control},
	author = {Camacho, J. and Parrilla, M. and Fritsch, C.},
	month = may,
	year = {2009},
	pages = {958--974},
}

@article{rodriguez-molares_generalized_2020,
	title = {The {Generalized} {Contrast}-to-{Noise} {Ratio}: {A} {Formal} {Definition} for {Lesion} {Detectability}},
	volume = {67},
	issn = {1525-8955},
	shorttitle = {The {Generalized} {Contrast}-to-{Noise} {Ratio}},
	doi = {10.1109/TUFFC.2019.2956855},
	number = {4},
	urldate = {2023-11-28},
	journal = {IEEE Transactions on Ultrasonics, Ferroelectrics, and Frequency Control},
	author = {Rodriguez-Molares, Alfonso and Rindal, Ole Marius Hoel and D’hooge, Jan and Måsøy, Svein-Erik and Austeng, Andreas and Lediju Bell, Muyinatu A. and Torp, Hans},
	month = apr,
	year = {2020},
	pages = {745--759},
}

@article{perrot_so_2021,
	title = {So you think you can {DAS}? {A} viewpoint on delay-and-sum beamforming},
	volume = {111},
	issn = {0041-624X},
	shorttitle = {So you think you can {DAS}?},
	doi = {10.1016/j.ultras.2020.106309},
	urldate = {2023-12-15},
	journal = {Ultrasonics},
	author = {Perrot, Vincent and Polichetti, Maxime and Varray, François and Garcia, Damien},
	month = mar,
	year = {2021},
	pages = {106309},
}

@article{wagner_statistics_1983,
	title = {Statistics of {Speckle} in {Ultrasound} {B}-{Scans}},
	volume = {30},
	issn = {2162-1403},
	doi = {10.1109/T-SU.1983.31404},
	number = {3},
	urldate = {2024-03-07},
	journal = {IEEE Transactions on Sonics and Ultrasonics},
	author = {Wagner, R.F. and Smith, S.W. and Sandrik, J.M. and Lopez, H.},
	month = may,
	year = {1983},
	pages = {156--163},
}

@article{gardner_grating_2024,
	title = {Grating lobe mitigation on large-pitch arrays using null subtraction imaging},
	volume = {140},
	issn = {0041-624X},
	doi = {10.1016/j.ultras.2024.107302},
	urldate = {2025-01-09},
	journal = {Ultrasonics},
	author = {Gardner, Mick and Miller, Rita J. and Oelze, Michael L.},
	month = may,
	year = {2024},
	pages = {107302},
}

@article{huang_scoliotic_2019,
	title = {Scoliotic {Imaging} {With} a {Novel} {Double}-{Sweep} 2.5-{Dimensional} {Extended} {Field}-of-{View} {Ultrasound}},
	volume = {66},
	issn = {1525-8955},
	doi = {10.1109/TUFFC.2019.2920422},
	number = {8},
	urldate = {2025-01-09},
	journal = {IEEE Transactions on Ultrasonics, Ferroelectrics, and Frequency Control},
	author = {Huang, Qinghua and Deng, Qifeng and Li, Le and Yang, Junlin and Li, Xuelong},
	month = aug,
	year = {2019},
	pages = {1304--1315},
}

@article{noorkoiv_assessment_2010,
	title = {Assessment of quadriceps muscle cross-sectional area by ultrasound extended-field-of-view imaging},
	volume = {109},
	issn = {1439-6327},
	doi = {10.1007/s00421-010-1402-1},
	language = {en},
	number = {4},
	urldate = {2025-01-09},
	journal = {European Journal of Applied Physiology},
	author = {Noorkoiv, M. and Nosaka, K. and Blazevich, A. J.},
	month = jul,
	year = {2010},
	pages = {631--639},
}

@article{yu_design_2020,
	title = {Design of a {Volumetric} {Imaging} {Sequence} {Using} a {Vantage}-256 {Ultrasound} {Research} {Platform} {Multiplexed} {With} a 1024-{Element} {Fully} {Sampled} {Matrix} {Array}},
	volume = {67},
	issn = {1525-8955},
	doi = {10.1109/TUFFC.2019.2942557},
	number = {2},
	urldate = {2025-01-09},
	journal = {IEEE Transactions on Ultrasonics, Ferroelectrics, and Frequency Control},
	author = {Yu, Jaesok and Yoon, Heechul and Khalifa, Yousuf M. and Emelianov, Stanislav Y.},
	month = feb,
	year = {2020},
	pages = {248--257},
}

@article{jensen_anatomic_2022,
	title = {Anatomic and {Functional} {Imaging} {Using} {Row}–{Column} {Arrays}},
	volume = {69},
	issn = {1525-8955},
	doi = {10.1109/TUFFC.2022.3191391},
	number = {10},
	urldate = {2025-01-09},
	journal = {IEEE Transactions on Ultrasonics, Ferroelectrics, and Frequency Control},
	author = {Jensen, Jørgen Arendt and Schou, Mikkel and Jørgensen, Lasse Thurmann and Tomov, Borislav G. and Stuart, Matthias Bo and Traberg, Marie Sand and Taghavi, Iman and Øygaard, Sigrid Huesebø and Ommen, Martin Lind and Steenberg, Kitty and Thomsen, Erik Vilain and Panduro, Nathalie Sarup and Nielsen, Michael Bachmann and Sørensen, Charlotte Mehlin},
	month = oct,
	year = {2022},
	pages = {2722--2738},
}

@article{poon_three-dimensional_2006,
	title = {Three-dimensional extended field-of-view ultrasound},
	volume = {32},
	issn = {0301-5629},
	doi = {10.1016/j.ultrasmedbio.2005.11.003},
	number = {3},
	urldate = {2025-02-06},
	journal = {Ultrasound in Medicine \& Biology},
	author = {Poon, Tony C. and Rohling, Robert N.},
	month = mar,
	year = {2006},
	pages = {357--369},
}

@article{treeby_rapid_2018,
	title = {Rapid calculation of acoustic fields from arbitrary continuous-wave sources},
	volume = {143},
	issn = {0001-4966},
	doi = {10.1121/1.5021245},
	number = {1},
	urldate = {2025-03-27},
	journal = {The Journal of the Acoustical Society of America},
	author = {Treeby, Bradley E. and Budisky, Jakub and Wise, Elliott S. and Jaros, Jiri and Cox, B. T.},
	month = jan,
	year = {2018},
	pages = {529--537},
}

@article{martin_simulating_2016,
	title = {Simulating {Focused} {Ultrasound} {Transducers} {Using} {Discrete} {Sources} on {Regular} {Cartesian} {Grids}},
	volume = {63},
	issn = {1525-8955},
	doi = {10.1109/TUFFC.2016.2600862},
	number = {10},
	urldate = {2025-03-27},
	journal = {IEEE Transactions on Ultrasonics, Ferroelectrics, and Frequency Control},
	author = {Martin, Eleanor and Ling, Yan To and Treeby, Bradley E.},
	month = oct,
	year = {2016},
	pages = {1535--1542},
}

@article{ramalli_design_2022,
	title = {Design, {Implementation}, and {Medical} {Applications} of 2-{D} {Ultrasound} {Sparse} {Arrays}},
	volume = {69},
	issn = {1525-8955},
	doi = {10.1109/TUFFC.2022.3162419},
	number = {10},
	urldate = {2025-07-22},
	journal = {IEEE Transactions on Ultrasonics, Ferroelectrics, and Frequency Control},
	author = {Ramalli, Alessandro and Boni, Enrico and Roux, Emmanuel and Liebgott, Hervé and Tortoli, Piero},
	month = oct,
	year = {2022},
	pages = {2739--2755},
}

@article{chavignon_3d_2022,
	title = {{3D} {Transcranial} {Ultrasound} {Localization} {Microscopy} in the {Rat} {Brain} {With} a {Multiplexed} {Matrix} {Probe}},
	volume = {69},
	issn = {1558-2531},
	doi = {10.1109/TBME.2021.3137265},
	number = {7},
	urldate = {2025-08-28},
	journal = {IEEE Transactions on Biomedical Engineering},
	author = {Chavignon, Arthur and Heiles, Baptiste and Hingot, Vincent and Orset, Cyrille and Vivien, Denis and Couture, Olivier},
	month = jul,
	year = {2022},
	pages = {2132--2142},
}

@inproceedings{savord_fully_2003,
	title = {Fully sampled matrix transducer for real time {3D} ultrasonic imaging},
	volume = {1},
	doi = {10.1109/ULTSYM.2003.1293556},
	urldate = {2025-09-03},
	booktitle = {{IEEE} {Symposium} on {Ultrasonics}, 2003},
	author = {Savord, B. and Solomon, R.},
	month = oct,
	year = {2003},
	pages = {945--953 Vol.1},
}

@article{castrignano_impact_2025,
	title = {On the {Impact} of {Microbeamformers} in 3-{D} {High} {Frame} {Rate} {Ultrasound} {Imaging}: {A} {Simulation} {Study}},
	volume = {72},
	issn = {1558-2531},
	shorttitle = {On the {Impact} of {Microbeamformers} in 3-{D} {High} {Frame} {Rate} {Ultrasound} {Imaging}},
	doi = {10.1109/TBME.2025.3529198},
	number = {6},
	urldate = {2025-09-03},
	journal = {IEEE Transactions on Biomedical Engineering},
	author = {Castrignano, Lorenzo and Tortoli, Piero and Matrone, Giulia and Crocco, Marco and Savoia, Alessandro Stuart and Ramalli, Alessandro},
	month = jun,
	year = {2025},
	pages = {1941--1950},
}

@article{favre_transcranial_2023,
	title = {Transcranial {3D} ultrasound localization microscopy using a large element matrix array with a multi-lens diffracting layer: an in vitro study},
	volume = {68},
	issn = {0031-9155},
	shorttitle = {Transcranial {3D} ultrasound localization microscopy using a large element matrix array with a multi-lens diffracting layer},
	doi = {10.1088/1361-6560/acbde3},
	language = {en},
	number = {7},
	urldate = {2025-09-04},
	journal = {Physics in Medicine \& Biology},
	author = {Favre, Hugues and Pernot, Mathieu and Tanter, Mickael and Papadacci, Clément},
	month = mar,
	year = {2023},
	pages = {075003},
}

@article{favre_boosting_2022,
	title = {Boosting transducer matrix sensitivity for {3D} large field ultrasound localization microscopy using a multi-lens diffracting layer: a simulation study},
	volume = {67},
	issn = {0031-9155},
	shorttitle = {Boosting transducer matrix sensitivity for {3D} large field ultrasound localization microscopy using a multi-lens diffracting layer},
	doi = {10.1088/1361-6560/ac5f72},
	language = {en},
	number = {8},
	urldate = {2025-09-04},
	journal = {Physics in Medicine \& Biology},
	author = {Favre, Hugues and Pernot, Mathieu and Tanter, Mickael and Papadacci, Clément},
	month = apr,
	year = {2022},
	pages = {085009},
}

@article{synnevag_adaptive_2007,
	title = {Adaptive {Beamforming} {Applied} to {Medical} {Ultrasound} {Imaging}},
	volume = {54},
	issn = {1525-8955},
	doi = {10.1109/TUFFC.2007.431},
	number = {8},
	urldate = {2025-10-03},
	journal = {IEEE Transactions on Ultrasonics, Ferroelectrics, and Frequency Control},
	author = {Synnevag, Johan Fredrik and Austeng, Andreas and Holm, Sverre},
	month = aug,
	year = {2007},
	pages = {1606--1613},
}

@article{montaldo_coherent_2009,
	title = {Coherent plane-wave compounding for very high frame rate ultrasonography and transient elastography},
	volume = {56},
	issn = {1525-8955},
	doi = {10.1109/TUFFC.2009.1067},
	number = {3},
	urldate = {2025-10-03},
	journal = {IEEE Transactions on Ultrasonics, Ferroelectrics, and Frequency Control},
	author = {Montaldo, Gabriel and Tanter, Mickaël and Bercoff, Jérémy and Benech, Nicolas and Fink, Mathias},
	month = mar,
	year = {2009},
	pages = {489--506},
}

@article{tsui_flexible_2012,
	title = {A flexible gel pad as an effective medium for scanning irregular surface anatomy},
	volume = {59},
	issn = {1496-8975},
	url = {https://doi.org/10.1007/s12630-011-9623-2},
	doi = {10.1007/s12630-011-9623-2},
	language = {en},
	number = {2},
	urldate = {2026-03-04},
	journal = {Canadian Journal of Anesthesia/Journal canadien d'anesthésie},
	author = {Tsui, Ban C. H. and Tsui, Jenkin},
	month = feb,
	year = {2012},
	pages = {226--227},
}

@article{corvino_utility_2020,
	title = {Utility of a gel stand-off pad in the detection of {Doppler} signal on focal nodular lesions of the skin},
	volume = {23},
	issn = {1876-7931},
	url = {https://doi.org/10.1007/s40477-019-00376-3},
	doi = {10.1007/s40477-019-00376-3},
	language = {en},
	number = {1},
	urldate = {2026-03-04},
	journal = {Journal of Ultrasound},
	author = {Corvino, Antonio and Sandomenico, Fabio and Corvino, Fabio and Campanino, Maria Raffaela and Verde, Francesco and Giurazza, Francesco and Tafuri, Domenico and Catalano, Orlando},
	month = mar,
	year = {2020},
	pages = {45--53},
}

@article{rathod_review_2020,
	title = {A {Review} of {Acoustic} {Impedance} {Matching} {Techniques} for {Piezoelectric} {Sensors} and {Transducers}},
	volume = {20},
	copyright = {http://creativecommons.org/licenses/by/3.0/},
	issn = {1424-8220},
	url = {https://www.mdpi.com/1424-8220/20/14/4051},
	doi = {10.3390/s20144051},
	language = {en},
	number = {14},
	urldate = {2026-03-12},
	journal = {Sensors},
	publisher = {Multidisciplinary Digital Publishing Institute},
	author = {Rathod, Vivek T.},
	month = jul,
	year = {2020},
}

@book{kinsler_fundamentals_2012,
	edition = {4th},
	title = {Fundamentals of {Acoustics}},
	isbn = {978-0-471-84789-2},
	author = {Kinsler, Lawrence and Frey, Austin and Coppens, Alan and Sanders, James},
	month = oct,
	year = {2012},
}

@article{afrakhteh_coherent_2021,
	title = {Coherent {Plane} {Wave} {Compounding} {Combined} {With} {Tensor} {Completion} {Applied} for {Ultrafast} {Imaging}},
	volume = {68},
	issn = {1525-8955},
	url = {https://ieeexplore.ieee.org/abstract/document/9448208},
	doi = {10.1109/TUFFC.2021.3087504},
	number = {10},
	urldate = {2026-04-14},
	journal = {IEEE Transactions on Ultrasonics, Ferroelectrics, and Frequency Control},
	author = {Afrakhteh, Sajjad and Behnam, Hamid},
	month = oct,
	year = {2021},
	pages = {3094--3103},
}

@article{afrakhteh_two-dimensional_2023,
	title = {A two-dimensional angular interpolation based on radial basis functions for high frame rate ultrafast imaging},
	volume = {154},
	issn = {0001-4966},
	url = {https://doi.org/10.1121/10.0022515},
	doi = {10.1121/10.0022515},
	number = {5},
	urldate = {2026-04-14},
	journal = {The Journal of the Acoustical Society of America},
	author = {Afrakhteh, Sajjad and Iacca, Giovanni and Demi, Libertario},
	month = nov,
	year = {2023},
	pages = {3454--3465},
}

\end{document}